\documentclass[lettersize,journal]{IEEEtran}
\usepackage{array}
\usepackage{textcomp}
\usepackage{stfloats}
\usepackage{url}
\usepackage{verbatim}
\usepackage{graphicx}
\usepackage{cite}

\usepackage{amsmath, bm, amsfonts}
\usepackage{epstopdf}
\usepackage{mathtools}
\usepackage{multirow}
\usepackage{algorithm}
\usepackage{algpseudocodex}

\newcommand{\argmin}{\mathop{\rm arg~min}\limits}
\newcommand{\bhline}[1]{\noalign{\hrule height #1}}
\usepackage{booktabs}

\makeatletter
\let\MYcaption\@makecaption
\makeatother

\usepackage[font=footnotesize]{subcaption}

\makeatletter
\let\@makecaption\MYcaption
\makeatother

\usepackage{subcaption}

\begin{document}

\title{Algorithm Unrolling-based Denoising of Multimodal Graph Signals}

\author{
    Hayate Kojima, 
    Keigo Takanami,~\IEEEmembership{Student member,~IEEE}, 
    Junya Hara, 
    Yukihiro Bandoh, 
    Seishi Takamura, \\
    Hiroshi Higashi,~\IEEEmembership{Member,~IEEE}, and 
    Yuichi Tanaka,~\IEEEmembership{Senior member,~IEEE}
    \thanks{H. Kojima and K. Takanami contributed equally for this work. The preliminary version of this paper is presented in \cite{takanamiMultimodadlGraphSignal2023b}. 
    H. Kojima, J. Hara, H. Higashi and Y. Tanaka are with the Graduate School of Engineering, The University of Osaka, Osaka, Japan. 
    K. Takanami is with the Graduate School of Engineering, Tokyo University of Agriculture and Technology, Tokyo, Japan. 
    Y. Bandoh is with Shimonoseki City University, Yamaguchi, Japan. 
    S. Takamura is with Graduate School of Computer and Information Sciences, Hosei University, Tokyo, Japan, and NTT Corporation, Kanagawa, Japan. 
    Corresponding author: Hayate Kojima (email: h-kojima@msp-lab.org).}
    }

\markboth{IEEE Transactions on Signal and Information Processing Over Networks,~Vol.~xx, 2026}%
{Kojima \MakeLowercase{\textit{et al.}}: Algorithm Unrolling-based Denoising of Multimodal Graph Signals}

\maketitle

\begin{abstract}
We propose a denoising method for multimodal graph signals by an alternating minimization scheme that sequentially solves signal restoration and graph learning problems.
Many complex-structured data, i.e., those on sensor networks, can capture multiple modalities at each measurement point, referred to as \textit{modalities}. 
They are also assumed to have an underlying structure or correlations in modality as well as space.
Such multimodal data are regarded as graph signals on a \textit{twofold graph} and they are often corrupted by noise.
Furthermore, their spatial/modality relationships are not always given a priori: We need to estimate twofold graphs during a denoising algorithm.
In this paper, we consider a signal denoising method on twofold graphs, where graphs are learned simultaneously.
Specifically, the graph learning subproblems are solved using the primal-dual splitting (PDS) algorithm, while the signal update has a closed-form solution.
Parameters in this iterative algorithm are learned from training data by unrolling the iteration with deep algorithm unrolling.
Experimental results on synthetic and real-world data demonstrate that the proposed method outperforms existing model- and deep learning-based graph signal denoising methods.
\end{abstract}

\begin{IEEEkeywords}
Multimodal data, signal denoising, graph learning, deep algorithm unrolling.
\end{IEEEkeywords}

\section{INTRODUCTION} \label{section:introduction}

\IEEEPARstart{S}{ensor} networks have been extensively utilized in various applications including environmental monitoring, Internet of Things (IoT), and home automation systems \cite{akyildizSurveySensorNetworks2002, akyildizWirelessSensorNetworks2002,alemdarWirelessSensorNetworks2010, tubaishatSensorNetworksOverview2003}.
Observed signals on sensor networks often corrupted by noise which affect to performances for the following applications.
Therefore, denoising is a fundamental task for sensor network-based signal processing systems \cite{krishnamurthi2020Overview, tay2021Sensor}.

Networks are mathematically represented by graphs and signals on a network are called \textit{graph signals}.
Graph signal processing (GSP) includes theory and algorithms to analyze graph signals \cite{ortega2018Graph, leus2023Graph, tanakaSamplingSignalsGraphs2020}.
GSP mainly considers \textit{spatially distributed} signal values whose domain is represented as nodes of a \textit{spatial graph}.

\begin{figure}
    \centering
    \includegraphics[width=0.95\hsize]{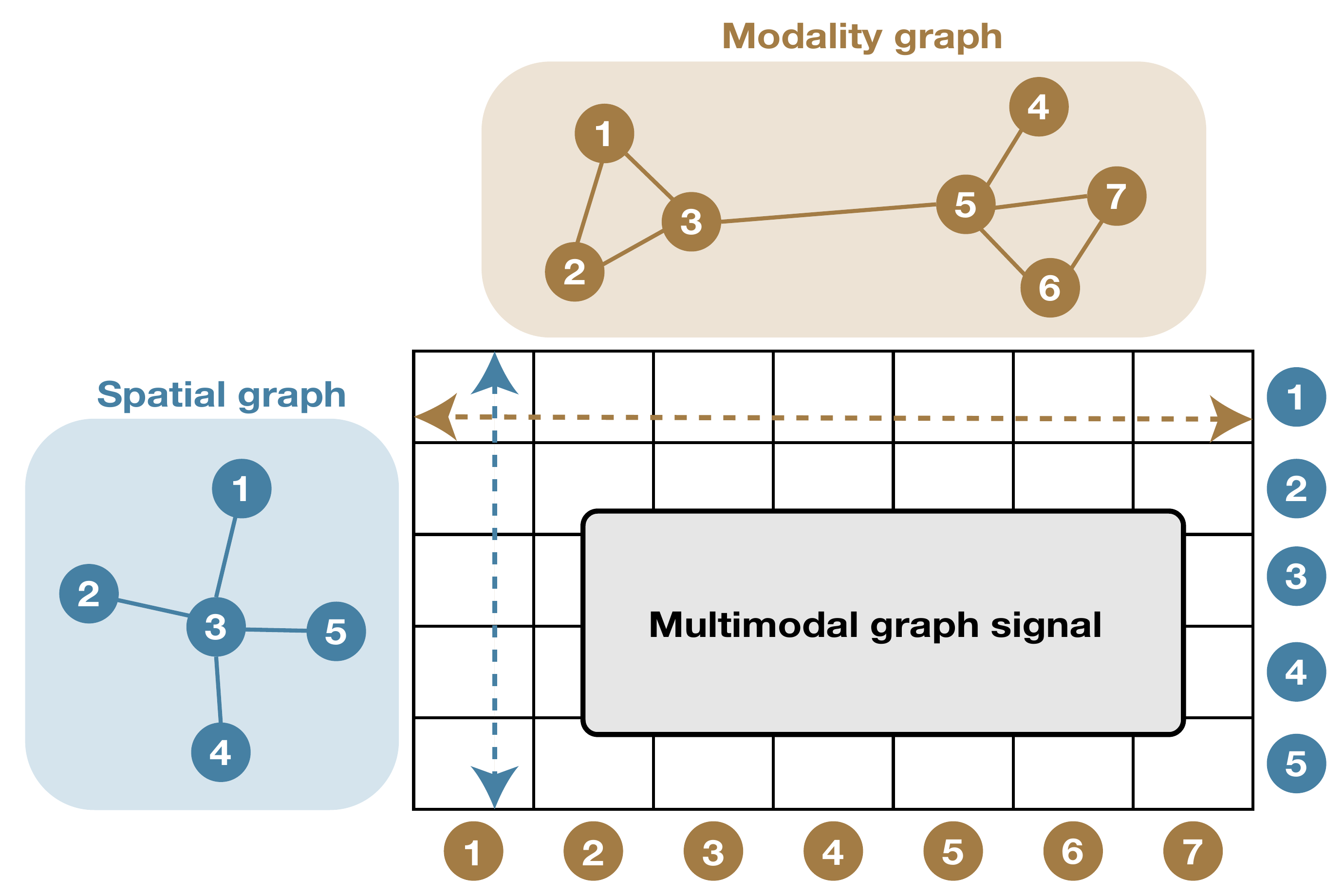}
    \caption{A multimodal graph signal on a twofold graph. }
    \label{figure:multimodal_graph_signal}
\end{figure}

In sensor networks, however, we also encounter relationships among \textit{modalities} since sensors often collect multiple features like temperature and humidity.
Some modalities are strongly correlated and the others are not.
This can be represented as a \textit{modality graph}.
It is visualized in Fig.~\ref{figure:multimodal_graph_signal}.
As a result, we can assume many sensor network data can be viewed as signals lying on a \textit{twofold graph} \cite{nagahama2022Multimodal}.

There are three main types of approaches to graph signal restoration: 1) model-based approach, 2) data-driven approach, and 3) hybrid approach.

The model-based approach designs a regularized objective function based on the prior for the observed data and finds its minimum solution.
The most widely used assumption is signal smoothness \cite{shumanEmergingFieldSignal2013, isufi2024Graph}.
It is often enforced via graph low-pass filters \cite{onukiGraphSignalDenoising2016, zhang2008Grapha}, Sobolev norms \cite{giraldo2022Reconstructiona}, or graph wavelets/filter banks \cite{watanabe2023Graph}. 
Piecewise constant signals have also been studied.
In this case, total variation (TV) is often considered \cite{Total_generalized_variation_for_graph_signals:Ono:ICASSP:2015}. 
Furthermore, low-rank approximations of the graph signal matrix are utilized to recover the underlying data structure, often in combination with smoothness constraints \cite{wu2025Robust, chen2015Signal}. 
More recently, generative approaches such as diffusion models have also been introduced to characterize complex signal distributions \cite{zhu2024Graph}. 
These model-based methods are typically solved using iterative algorithms. 
Note that the model-based method usually needs to tune hyperparameters,
such as the regularization strengths.
They are fixed
throughout the iterative process, and in many cases, they are manually tuned.

The data-driven approach learns (hyper)parameters from given training data with machine learning techniques.
The current majority is based on graph neural networks (GNNs) \cite{liang2022Survey,ma2025Acceleration}.
The representative example is graph convolutional networks (GCNs) 
\cite{DBLP:conf/iclr/KipfW17,bhatti2023Deep,chen2024Survey,shen2022GCNDenoiser,yan2024GNFNet}.
However, GNNs do not always outperform model-based methods for signal restoration \cite{jian2024Kernela}.
This may stem from the fact that, in contrast to uniformly sampled signals like image and audio signals, deep layers for GNNs may not result in good performance \cite{khemani2024Reviewa,chen2021Grapha}.
We also encounter a challenge for training data: We need to collect enough number of data for training, but this is not often the case for the graph setting.

The hybrid approach has been proposed to overcome the aforementioned challenges in the model- and data-driven methods.
This often integrates (sometimes black-box) denoisers in an iterative algorithm \cite{Interpolation_and_Denoising_of_Graph_Signals_Using_Plug-and-play_ADMM:Yazaki:ICASSP:2019}.
In terms of the hyperparameter tuning, deep algorithm unrolling (DAU)
\cite{Learning_fast_approximations_of_sparse_coding:DUA:ICML:2010,monga2021Algorithm} has attracted the signal processing community.
DAU organizes each iteration of the model-based approach as a layer of neural network,
and learns the hyperparameters in each layer from training data.
DAU has the following benefits.
First, it automatically learns hyperparameters in the iterative algorithm.
While the convexity of the whole minimization problem is not guaranteed in general, it practically works well due to the different tuned parameters in different layers.
Second, DAU usually requires a smaller number of parameters than the \textit{full} neural networks: This contributes to reducing the training data.
Third, since it is based on iterative minimization algorithms, the algorithm maintains some sense of interpretablity.

For graph signal restoration, several DAU-based methods for \textit{single modal graph signals} have been proposed
 \cite{chen2021Grapha, Graph_Signal_Restoration:Nagahama:TSP:2022}.
An extension for multimodal graph signals has also been proposed 
\cite{nagahama2022Multimodal, yang2022GraphBased}, leveraging two graphs representing spatial and modality relationships.
However, all of them assume the graphs are given a priori.

In this paper, we consider a denoising problem of multimodal signals on a twofold graph
under the condition that the spatial/modality graphs are not given a priori: We need to learn/estimate twofold graphs for that case.
The proposed method simultaneously learns a twofold graph from observed data during denoising.
Specifically, an iterative algorithm is performed alternately for signal denoising and graph learning.

In the proposed method, first, we design an optimization problem of denoising on twofold graphs and derive an iterative algorithm that leverages the smoothness of signals on the twofold graph, i.e., assuming the signals are smooth both on a spatial graph and a modality graph.
Each iteration consists of two building blocks: Denoising in the spatial domain and that in the modality domain. 
Each building block is further divided into two phases: Graph learning and signal smoothing.
In the graph learning phase, we utilize an existing technique \cite{kalofoliasHowLearnGraph2016, dongLearningLaplacianMatrix2016} which assumes signal smoothness on the graph in the corresponding domain.
In the signal smoothing phase, we apply the well-known Tikhonov regularization \cite{shumanEmergingFieldSignal2013} to the multimodal signals.
Finally, we \textit{unroll} the algorithm using deep algorithm unrolling (DAU) \cite{monga2021Algorithm}.
The parameters in each iteration are learned from the training data.

In numerical experiments for synthetic and real-world signals, the proposed method exhibits lower root-mean-squared errors (RMSEs) than existing model-based, GCN-based and DAU-based denoising methods.

The remainder of this paper is organized as follows: 
Section \ref{section:related_work} reviews related work on graph signal denoising and graph learning. 
Section \ref{section:proposed} presents the proposed multimodal graph signal denoising framework, including the problem formulation and the hyperparameter optimization approach based on deep algorithm unrolling (DAU). 
Section \ref{section:exp} describes the experimental setup and reports the results on both synthetic and real-world datasets, demonstrating the effectiveness of the proposed method. 
Finally, Section \ref{section:conclusion} concludes the paper and outlines potential directions for future work.

\textit{Notation: } 
The notation of this paper is summarized in Table.~\ref{table:notation}.
We consider a weighted undirected graph $\mathcal{G}=(\mathcal{V}, \mathcal{E})$, where $\mathcal{V} = \{ v_{0}, \ldots, v_{N-1} \}$ and $\mathcal{E} \subseteq \mathcal{V} \times \mathcal{V}$ represent sets of nodes and edges, respectively.
Unless otherwise specified, $N = |\mathcal{V}|$ represents the number of nodes.
The adjacency matrix of $\mathcal{G}$ is denoted by $\mathbf{W}$, where the $(i,j)$-element of $\mathbf{W}$, $w_{i,j} \geq 0$, is the edge weight between the $i$th and $j$th nodes; $w_{i,j}=0$ for unconnected nodes.
The degree matrix $\mathbf{D}$ is defined as $\mathbf{D}=\textrm{diag}(d_{0}, \ldots, d_{N-1})$, where $d_{i}=\sum_{j} w_{i,j}$ is the $i$th diagonal element.
We use graph Laplacian $\mathbf{L}:= \mathbf{D} - \mathbf{W}$ as a graph operator.
We can arbitrarily assign indices of edges as 
$\mathcal{E}=\{e_{q}\}_{q=0}^{|\mathcal{E}|-1}$. 
The weighted graph incidence matrix of $\mathcal{G}$ is defined as $\mathbf{M} \in \mathbb{R}^{N \times |\mathcal{E}|}$, where the $(p,q)$-element of $\mathbf{M}$ is defined as follows:
\[
    [\mathbf{M}]_{p,q} = 
    \begin{dcases}
         \sqrt{w_{i,j}} & \textrm{if $p=i$ and $e_{q}=(v_i,v_j)$}  \\
        -\sqrt{w_{i,j}} & \textrm{if $p=j$ and $e_{q}=(v_i,v_j)$}  \\
        0 & \textrm{otherwise}.
    \end{dcases}
\]
A graph signal $\mathbf{x} \in \mathbb{R}^{N}$ is defined as a mapping from the node set to the set of real numbers, i.e., $x_{i}:\mathcal{V} \rightarrow \mathbb{R}$. 
We assume that $\mathbf{x}$ is smooth on $\mathcal{G}$. The smoothness of $\mathbf{x}$ is measured using the Laplacian quadratic form $\mathbf{x}^\top\mathbf{L}\mathbf{x}$. If it is small, $\mathbf{x}$ is considered to be smooth on $\mathcal{G}$.
The graph Fourier transform (GFT) of $\mathbf{x}$ is defined as $\hat{\mathbf{x}} = \mathbf{U}^{\top} \mathbf{x}$ where the GFT matrix $\mathbf{U}$ is obtained by the eigendecomposition of the graph Laplacian $\mathbf{L} = \mathbf{U \Lambda} \mathbf{U}^{\top}$ with the eigenvalue matrix $\mathbf{\Lambda} = \textrm{diag}(\lambda_{0}, \ldots, \lambda_{N-1})$.
We refer to $\lambda_{i}$ as a \textit{graph frequency}.

The domain index is denoted by $e=\{s,m\}$, where $s$ and $m$ correspond to the spatial and modality domains, respectively.
In this manner, spatial and modality graphs are denoted by $\mathcal{G}_{s}=(\mathcal{V}_{s}, \mathcal{E}_{s},\mathbf{W}_{s})$ and $\mathcal{G}_{m}=(\mathcal{V}_{m}, \mathcal{E}_{m},\mathbf{W}_{m})$, respectively.
$N_{s} = |\mathcal{V}_{s}|$ and $N_{m} = |\mathcal{V}_{m}|$ represents the number of nodes of each graph.
A multimodal graph signal is defined as $\mathbf{X} \in \mathbb{R}^{N_{s} \times N_{m}}$ (see Fig.~\ref{figure:multimodal_graph_signal}).

\begin{table}[t]
    \caption{Symbols used in this paper.}
    \label{table:notation}
    \centering
    \begin{tabular}{c|c} \bhline{1.1pt}
        Symbols & Definitions \\ \hline \hline
        $x, \mathbf{x}, \mathbf{X}$                            & scalar, vector, and matrix \\ 
        $x_{i}, [\mathbf{x}]_{i}$                      & $i$-th element of $\mathbf{x}$\\
        $[\mathbf{X}]_{i,j}$                           & $(i,j)$-th element of $\mathbf{X}$ \\
        $[\mathbf{X}]_{i,:}$              & $i$-th row vector of $\mathbf{X}$ \\
        $[\mathbf{X}]_{:,j}$                 & $j$-th column vector of $\mathbf{X}$ \\
        $\mathbf{1}_{N}$                              & all-ones vector with the length $N$ \\
        $\mathbf{I}_{N}$                                & identity matrix with the size $N \times N$ \\ \hline
        $\textrm{log}(\mathbf{x})$ & vector with the logarithm of each element of $\mathbf{x}$  \\
        $\textrm{diag}(x_{0}, \ldots, x_{N-1})$ & diagonal matrix whose $(i,i)$-elements are $x_{i}$ \\
        $\textrm{Diag}(\mathbf{X})$ & vector whose $i$-th elements are $[\mathbf{X}]_{i,i}$ \\
        $\textrm{tr}(\mathbf{X})$ & trace of $\mathbf{X}$ \\
        $\|\mathbf{x}\|_{p}$  & $\ell_{p}$ norm of $\mathbf{x}$ \\
        $\|\mathbf{X}\|_{F}$ & Frobenius norm of $\mathbf{X}$ \\
        $\mathbf{X}^{\dagger}$   & Moore-Penrose pseudoinverse of $\mathbf{X}$ \\
        $\mathbf{X} \circ \mathbf{Y}$ & Hadamard product of $\mathbf{X}$ and $\mathbf{Y}$ \\ \hline
    \end{tabular}
\end{table}

\section{RELATED WORK} \label{section:related_work}
In this section, we review graph signal denoising and graph learning relevant to the formulation of the proposed method which is discussed below in Section \ref{section:proposed}.
First, We introduce existing model-based graph signal denoising methods \cite{shumanEmergingFieldSignal2013, nagahama2022Multimodal}.
Then, representative graph learning approaches are shown \cite{dongLearningGraphsData2019, kalofoliasHowLearnGraph2016}.

\subsection{GRAPH SIGNAL DENOISING} \label{subsection:denoising}

In a typical denoising setting, the following observation model is assumed:
\begin{equation}
    \mathbf{y} = \mathbf{x} + \mathbf{n},
    \label{eq:single_modal_observation}
\end{equation}
where $\mathbf{y} \in \mathbb{R}^{N}$ is the noisy observation, $\mathbf{x} \in \mathbb{R}^{N}$ is the unknown ground-truth signal, and $\mathbf{n} \in \mathbb{R}^{N}$ is additive white Gaussian noise (AWGN).

Signal denoising aims to recover $\mathbf{x}$ from $\mathbf{y}$ based on the assumption(s) on signal priors.
Similar to time- and spatial-domain signal processing, we assume smoothness of the signal on the graph $\mathcal{G}$, where the smoothness is measured by the Laplacian quadratic form $\mathbf{x}^{\top} \mathbf{L} \mathbf{x} = \|\mathbf{L}^{1/2} \mathbf{x}\|_2^2$.
Therefore, the well-known denoising with a Tikhonov regularization term is given as follows:
\begin{equation} \label{equation:tikhonov}
    \min_{\mathbf{x}} \| \mathbf{y} - \mathbf{x} \|_{2}^{2} + \alpha \mathbf{x}^{\top} \mathbf{L} \mathbf{x}, 
\end{equation}
where $\alpha>0$ is the regularization parameter.
Note that, the regularization term acts as a graph high-pass filter since $\mathbf{L}^{1/2} \mathbf{x} = \mathbf{U} \hat{g}(\bm{\Lambda}) \mathbf{U}^\top \mathbf{x}$ where the graph filter $\hat{g}(\lambda) = \lambda^{1/2}$.
As a result,
minimizing \eqref{equation:tikhonov} promotes smoothness of graph signals.

We can easily obtain the solution of \eqref{equation:tikhonov} in a closed form:
\begin{equation} \label{equation:filtering_in_graph_frequency}
    \widetilde{\mathbf{x}} = (\mathbf{I}_{N} + \alpha \mathbf{L})^{-1} \mathbf{y} = \mathbf{U} (\mathbf{I}_{N} + \alpha \mathbf{\Lambda})^{-1} \mathbf{U}^{\top} \mathbf{y}.
\end{equation}
Denoting by $\hat{h}(\mathbf{\Lambda}) = (\mathbf{I} + \alpha \mathbf{\Lambda})^{-1}$, \eqref{equation:filtering_in_graph_frequency} can be interpreted as a low-pass filter with the graph spectral kernel $\hat{g}(\lambda) = 1 / (1 + \alpha \lambda)$.

Beyond the Tikhonov regularization, a variety of filter kernels has been considered \cite{shumanEmergingFieldSignal2013, tomasiBilateralFilteringGray1998, onukiGraphSignalDenoising2016, gavishOptimalHardThreshold2014, zhang2008Grapha}.
Note that any smoothing filter can be applied to our framework as long as it fits our formulation.

The above-mentioned observation model in \eqref{eq:single_modal_observation} is the single modal case where each node is associated with a scalar.
In the multimodal setting, we easily extend the signal observation model as follows:
\begin{equation} \label{equation:definition_multimodal_graph_signal}
    \mathbf{Y} = \mathbf{X} + \mathbf{N},
\end{equation}
where $\mathbf{X}$, $\mathbf{Y}$, $\mathbf{N} \in \mathbb{R}^{N_{s} \times N_{m}}$ are the unknown ground-truth signal, noisy measurement, and AWGN, respectively.
In this case, the $(i,k)$ element of a matrix corresponds to the signal value at node $i$ and modal $k$.

Similar to the single-modal case, multimodal signal denoising aims to recover $\mathbf{X}$ from $\mathbf{Y}$ by utilizing signal priors.
In \cite{nagahama2022Multimodal}, multimodal graph signal denoising is introduced similar to the single modal case under the given twofold graph.
However, in a practical case, the twofold graph is not given, both for the spatial and modality graphs.

\subsection{GRAPH LEARNING} \label{subsection:graph_learning}
Many GSP techniques assume graph Laplacian is given a priori. 
However, this is not often the case: We need to estimate/learn graphs from prior knowledge or given data.
The classical method is $k$-nearest neighbor which has still been widely used \cite{mateos2019Connecting}.
However, the obtained graph does not always reflect the statistics of observed signals.

As a more data-driven graph estimation approach, graph learning has been widely studied in GSP\cite{dongLearningLaplacianMatrix2016,kalofoliasHowLearnGraph2016,DBLP:conf/iclr/KalofoliasP19,yamadaTimevaryingGraphLearning2019}.
Let $\mathbf{X}_{\textrm{single}} = \begin{bmatrix}\mathbf{x}_{0}, \dots, \mathbf{x}_{K-1}\end{bmatrix}\in \mathbb{R}^{N \times K}$ be a collection of $K$ single modal graph signals on a spatial graph.
Note that the column does not refer to modality in contrast to \eqref{equation:definition_multimodal_graph_signal}.

A typical goal of graph learning is to estimate a graph Laplacian $\mathbf{L}$ from $\mathbf{X}_{\textrm{single}}$.
Assuming the signal smoothness, a graph learning problem is often given
as follows:
\begin{equation} \label{equation:min_L}
    \min_{\mathbf{L} \in \mathcal{L}} \textrm{tr}(\mathbf{X}_{\textrm{single}}^{\top} \mathbf{L} \mathbf{X}_{\textrm{single}}) + f_{L}(\mathbf{L}),
\end{equation}
where $\mathcal{L} = \{ \mathbf{L} \in \mathbb{R}^{N \times N} ~|~ \mathbf{L} \mathbf{1}_{N} = \mathbf{0}, ~[\mathbf{L}]_{i,j} = [\mathbf{L}]_{j,i} \leq 0 ~(i \neq j) \}$ is the feasible set of graph Laplacians and $f_L(\mathbf{L})$ is a regularization function with respect to $\mathbf{L}$.
Since the first term in \eqref{equation:min_L} is nothing but the Laplacian quadratic form, it promotes the smoothness of signals on the graph.
Many regularization functions $f_L(\mathbf{L})$ have been proposed in literature: For example,
$f_L(\mathbf{L}) = \gamma \| \mathbf{L} \|_{F}^{2}$ where $\gamma>0$ is the regularization parameter in \cite{dongLearningLaplacianMatrix2016}. 
Constraints on time-varying graph learning has also been proposed in \cite{yamadaTimevaryingGraphLearning2019}.

A similar expression of graph learning to \eqref{equation:min_L} based on adjacency matrix has been proposed in \cite{kalofoliasHowLearnGraph2016}.
It utilizes a distance matrix $\mathbf{Z} \in \mathbb{R}^{N \times N}$ whose element is given by
\begin{equation} \label{equation:definition_pairwise_distance_matrix}
    [\mathbf{Z}]_{i,j} = \| [\mathbf{X}_{\textrm{single}}]_{i,:} - [\mathbf{X}_{\textrm{single}}]_{j,:} \|_{2}^{2}.
\end{equation}
We can then rewrite \eqref{equation:min_L} with $\mathbf{Z}$ as follows:
\begin{equation}
    \min_{\mathbf{W} \in \mathcal{W}} \|\mathbf{W} \circ \mathbf{Z}\|_1+ f_W(\mathbf{W}),
    \label{equation:min_W}
\end{equation}
where $\mathcal{W}= \{\mathbf{W}\in\mathbb{R}^{N\times N}_{\ge 0} ~|~ \mathbf{W}=\mathbf{W}^{\top}, [\mathbf{W}]_{i,i}=0\}$ is the feasible set of adjacency matrices and $f_W(\mathbf{W})$ is an arbitrary regularization function with respect to $\mathbf{W}$.

The first term in \eqref{equation:min_W} is equivalent to $2\textrm{tr}(\mathbf{X}_{\textrm{single}}^{\top}\mathbf{L} \mathbf{X}_{\textrm{single}})$ and it corresponds to the signal smoothness term in \eqref{equation:min_L}.
Also, 
many versions of $f_W(\mathbf{W})$ have been presented (see \cite{dongLearningGraphsData2019} and references therein):
For example, $f_W(\mathbf{W})$ is defined as follows in \cite{kalofoliasHowLearnGraph2016}:
\begin{equation}
  f_W(\mathbf{W}) = - \beta \mathbf{1}_{N}^{\top} \log(\mathbf{W} \mathbf{1}_{N}) + \frac{\gamma}{2} \|\mathbf{W} \|_{F}^{2},
  \label{equation:objective_kalofolias}
\end{equation}
where $\beta>0$ and $\gamma>0$ are parameters. The first term controls the connectivity of the graph, where a
larger $\beta$ promotes dense connections. The
second term regularizes the edge weights, where a larger $\gamma$ results in smaller edge weights.

Note that, while \cite{dongLearningLaplacianMatrix2016} considers a joint problem of graph learning and smoothing signals, multimodal versions with twofold graphs have not been presented so far.
In the following section, we introduce our proposed method by simultaneously considering signal smoothing and twofold graph learning.

\section{MULTIMODAL GRAPH SIGNAL DENOISING WITH SIMULTANEOUS GRAPH LEARNING} \label{section:proposed}
\begin{figure}
    \centering
    \includegraphics[width=0.95\hsize]{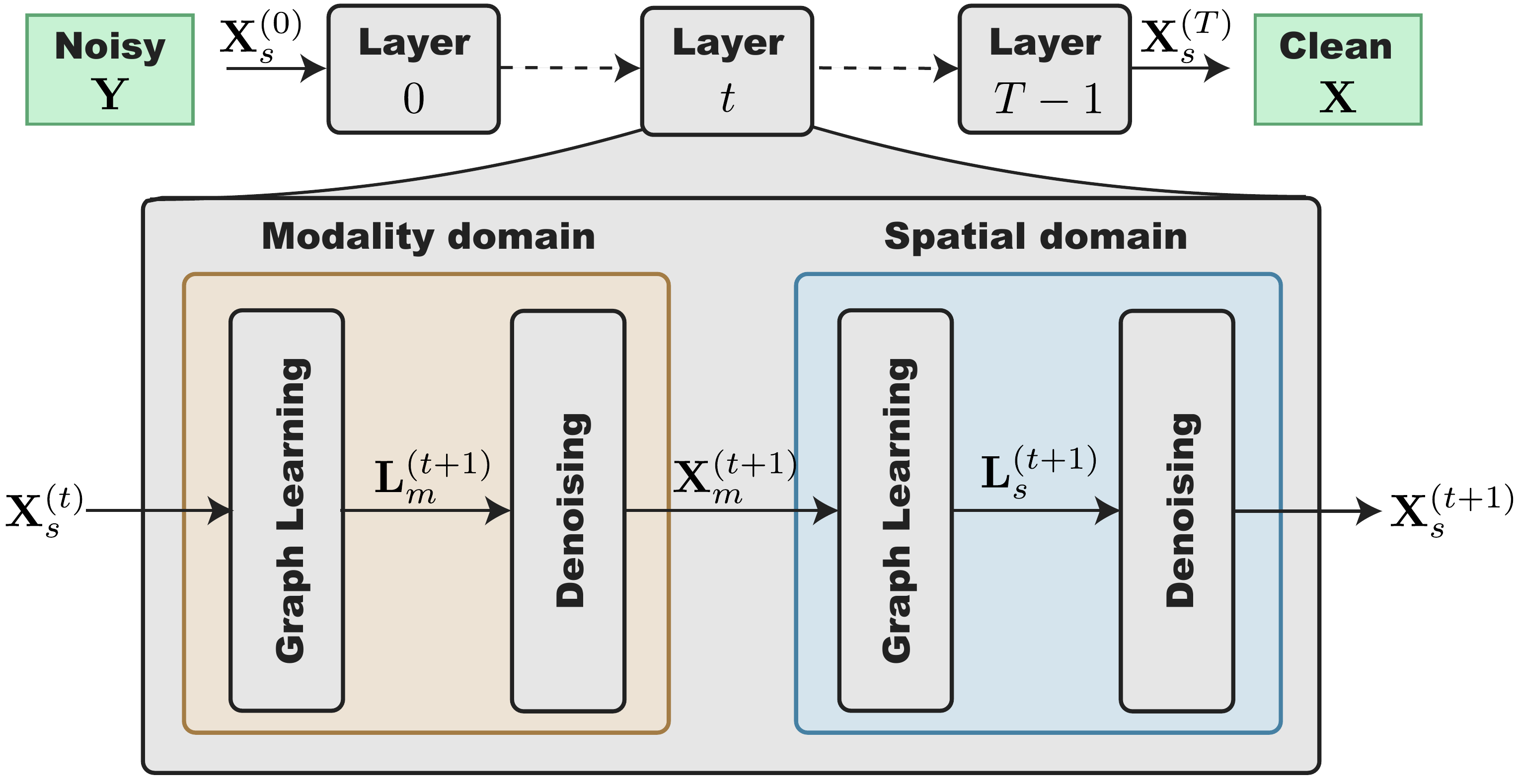}
    \caption{Overview of the proposed architecture.}
    \label{figure:proposed_networks}
\end{figure}

As mentioned above, existing graph signal denoising methods assume the underlying graph is given prior to denoising.
They also assume signals are single-modal.
However, they are not often the case.
In this section, we propose a denoising method for multimodal graph signals by learning twofold graphs simultaneously.
The overview of our method is illustrated in Fig.~\ref{figure:proposed_networks}.

\subsection{PROBLEM FORMULATION} \label{subsection:proposed-formulation}
We consider the measurement model $\mathbf{Y}=\mathbf{X}+\mathbf{N}$ in \eqref{equation:definition_multimodal_graph_signal}.
We aim to restore $\mathbf{X}$ from $\mathbf{Y}$ by simultaneously learning a spatial graph $\mathcal{G}_{s}$ and a modality graph $\mathcal{G}_{m}$.

Here, we assume the joint smoothness on the twofold graph: Each column of $\mathbf{X}$ is smooth on $\mathcal{G}_{s}$ and each row of $\mathbf{X}$ is smooth on $\mathcal{G}_{m}$.
Based on the smoothness assumption, we naturally extend the denoising for the single modal case in \eqref{equation:tikhonov} to the multimodal case as follows:
\begin{equation}
    \min_{\mathbf{X}, \mathbf{L}_{s} \in \mathcal{L}_{s}, \mathbf{L}_{m} \in \mathcal{L}_{m}}
    ~ \| \mathbf{Y} - \mathbf{X} \|_{F}^{2} + f_{s}(\mathbf{X}, \mathbf{L}_{s}) + f_{m}(\mathbf{X}^{\top}, \mathbf{L}_{m}),
    \label{equation:objective_proposed}
\end{equation}
where $\mathcal{L}_{e} = \{ \mathbf{L}_{e} \in \mathbb{R}^{N_{e} \times N_{e}} ~|~ \mathbf{L}_{e} \mathbf{1}_{N_{e}} = \mathbf{0}, ~[\mathbf{L}_{e}]_{i,j} = [\mathbf{L}_{e}]_{j,i} \leq 0 ~(i \neq j)\}$ is the set of valid graph Laplacians  for $e\in \{s,m\}$ and $f_{e}(\cdot)$ is a function with respect to multimodal graph signals $\mathbf{X}$ and graph Laplacian $\mathbf{L}_{e}$.

Since we consider a joint problem of signal smoothing and graph learning, the regularization term $f_{e}(\cdot)$ should reflect the both constraints.
In this paper, we consider the following $f_{e}(\cdot)$ as an extension of that used in \cite{dongLearningLaplacianMatrix2016,kalofoliasHowLearnGraph2016}:
\begin{equation}
    \begin{aligned}
    {f}_{e}(\mathbf{X}_{e},\mathbf{L}_{e}) 
    = 
    ~ &\alpha_{e} \textrm{tr}(\mathbf{X}_{e}^{\top} \mathbf{L}_{e} \mathbf{X}_{e}) - \beta_{e} \mathbf{1}_{N_{e}}^{\top} \log(\textrm{diag}(\mathbf{L}_{e}))\\
    & + \frac{\gamma_{e}}{2} \| \bm{\Omega}_{e} \circ \mathbf{L}_{e} \|_{F}^{2},
    \end{aligned}
    \label{equation:proposed_matrix_f}
\end{equation}
where $\alpha_{e}$, $\beta_{e}$, and $\gamma_{e}$ are positive parameters, and $\bm{\Omega}_{e} \in \{0,1\}^{N_{e} \times N_{e}}$ is a matrix which extracts off-diagonal elements of $\mathbf{L}_{e}$.
The first, second, and third terms in \eqref{equation:proposed_matrix_f} promote signal smoothness on the graph, node connectivity, and sparseness of edge weights.
This formulation promotes sparse graphs, which lead to computational efficiency in subsequent graph filtering steps. 
Note that our framework is flexible: Any differentiable graph learning method can be employed instead of the above one.

The optimization problem \eqref{equation:objective_proposed} with \eqref{equation:proposed_matrix_f} is jointly nonconvex. In this paper, we split it into the following convex optimization problems for graph signals ($\mathbf{X}_{e}$) and graph Laplacians ($\mathbf{L}_{e}$) as follows:
\begin{subequations}
    \begin{align}
        &\min_{\mathbf{X}_{m}} \| \mathbf{Y}^{\top} - \mathbf{X}_{m} \|_{F}^{2} + \alpha_{m} \textrm{tr}(\mathbf{X}_{m}^{\top} \mathbf{L}_{m} \mathbf{X}_{m}), \label{subequation:objective_function_Xm} \\
        &\min_{\mathbf{X}_{s}} \| \mathbf{X}_{m}^{\top} - \mathbf{X}_{s} \|_{F}^{2} + \alpha_{s} \textrm{tr}(\mathbf{X}_{s}^{\top} \mathbf{L}_{s} \mathbf{X}_{s}) \label{subequation:objective_function_Xs}, \\
        &\min_{\mathbf{L}_{m} \in \mathcal{L}_{m}} f_{m}(\mathbf{X}_{m}, \mathbf{L}_{m}) \label{subequation:objective_function_Lm}, \\
        &\min_{\mathbf{L}_{s} \in \mathcal{L}_{s}} f_{s}(\mathbf{X}_{s}, \mathbf{L}_{s}).
        \label{subequation:objective_function_Ls}
    \end{align}
    \label{equation:splitted_optimization}
\end{subequations}
As mentioned in Section \ref{section:proposed}, \eqref{subequation:objective_function_Xm} and \eqref{subequation:objective_function_Xs} have the closed form solutions.
Later, we also introduce that \eqref{subequation:objective_function_Lm} and \eqref{subequation:objective_function_Ls} can be solved by an iterative algorithm using primal-dual splitting (PDS) \cite{komodakisPlayingDualityOverview2015a} after appropriate transformations.
Therefore, if the parameters $\{\alpha_{e}, \beta_{e}, \gamma_{e}\}$ for $e\in \{s,m\}$ are fixed, each of the subproblems is convex.
Note that we train the parameters via DAU from the training data (Section \ref{section:proposed}-\ref{subsection:proposed_DAU}) to promote fewer iterations.

\subsection{Internal Iterative Algorithm}
Here, we provide the solutions to \eqref{subequation:objective_function_Lm} and \eqref{subequation:objective_function_Ls} using PDS algorithm.
In this paper, we consider undirected graphs without self-loops.
This means we can distinguish the graph Laplacian with its (lower or upper) triangular components $\bm{\ell}_{e} \in \mathbb{R}^{N_{e}(N_{e}-1)/2}$.
Let us define $\textrm{vec}(\mathbf{L}_{e}) \in \mathbb{R}^{N_{e}^{2}}$ as the vectorized version of $\mathbf{L}_{e}$.
By using $\bm{\ell}_{e}$, $\mathbf{L}_{e}$ can be represented as follows:
\begin{equation}
    \textrm{vec}(\mathbf{L}_{e}) = \bm{\Phi}_{e} \bm{\ell}_{e},
    \label{eq:vecLe}
\end{equation}
where $\bm{\Phi}_{e} \in \{-1,0,1\}^{N_{e}^{2} \times \frac{N_{e}(N_{e}-1)}{2}}$ is the duplication matrix \cite{abadirMatrixAlgebra2005}.
By using the relationship in \eqref{eq:vecLe}, each term in \eqref{subequation:objective_function_Lm} and \eqref{subequation:objective_function_Lm} can be rewritten as follows:
\begin{equation}
    \begin{aligned}
        \textrm{tr}(\mathbf{X}_{e}^{\top}\mathbf{L}_{e}\mathbf{X}_{e}) &= \textrm{vec}(\mathbf{X}_{e}\mathbf{X}_{e}^{\top})^{\top}\bm{\Phi}_{e} \bm{\ell}_{e}, \\
        \mathbf{1}_{N_{e}}^{\top}\textrm{log}(\textrm{diag}(\textbf{L}_{e}))
        &= \mathbf{1}_{N_{e}}^{\top}\textrm{log}(\bm{\Psi}_{e} \bm{\ell}_{e}), \\
        \| \bm{\Omega}_{e} \circ \mathbf{L}_{e} \|_{F}^{2} 
        &= 2 \|\bm{\ell}_{e}\|_{2}^{2},
    \end{aligned}
\end{equation}
where $\bm{\Psi}_{e} \in \{-1, 0\}^{N_{e} \times \frac{N_{e}(N_{e}-1)}{2}}$ is the matrix of a linear operator that satisfies $\textrm{diag}(\mathbf{L}_{e}) = \bm{\Psi}_{e} \bm{\ell}_{e}$.
As a result, we can rewrite \eqref{equation:proposed_matrix_f}
as follows:
\begin{equation}
    \begin{aligned}
        \min_{\bm{\ell}_{e}}
        ~& \alpha_{e} \textrm{vec}(\mathbf{X}_{e}\mathbf{X}_{e}^{\top})^{\top} \bm{\Phi}_{e}\bm{\ell}_{e} 
        - \beta_{e} \mathbf{1}_{N_{e}}^{\top} \textrm{log}(\bm{\Psi}_{e}\bm{\ell}_{e} ) \\
        &+ \gamma_{e} \| \bm{\ell}_{e} \|_{2}^{2}
        + \iota_{\cdot \leq \mathbf{0}}(\bm{\ell}_{e}),
    \end{aligned}
    \label{equation:objective_function_sub_vech_indicator}
\end{equation}
where the last term expresses the constraint of the graph Laplacian using the indicator function $\iota_{\cdot\leq\mathbf{0}}$, which is defined by
\begin{equation}
    \iota_{\cdot \leq \mathbf{0}} (\bm{\ell}_{e}) = 
    \begin{dcases}
        0      & \textrm{if } \ell_{i} \leq 0 \\
        \infty & \textrm{otherwise}. \\
    \end{dcases}
\end{equation}

In fact, \eqref{equation:objective_function_sub_vech_indicator} can be solved iteratively with the PDS algorithm (detailed derivations are shown in Appendix).
Its algorithm is shown in Algorithm \ref{algorithm:pds_form_e} where we omit the index $e$ for simplicity.

\subsection{Deep Algorithm Unrolling} \label{subsection:proposed_DAU}
We utilize DAU for determine hyperparameters in Algorithm \ref{algorithm:pds_form_e} to make the number of iterations smaller.
We replace six hyperparameters $\{\alpha_{e}, \beta_{e}, \gamma_{e}\}$ for $e\in \{s,m\}$ in \eqref{equation:proposed_matrix_f} with trainable ones defined as $ \bm{\Theta} \coloneqq \{\alpha_{e}^{(t)}$, $\beta_{e}^{(t)}$, $\gamma_{e}^{(t)}\}_{t=0}^{T-1}$, where $T$ is the number of iterations (layers).
This DAU-based approach is expected to learn appropriate parameters even when the training signals do not strictly satisfy the model-based assumptions (e.g., smoothness).
We summarize the proposed denoising algorithm in Algorithm \ref{algorithm:MGSD-LLap-DAU}.

Our unrolled algorithm is trained to minimize the mean squared error (MSE) of multimodal signals as follows:
\begin{equation}
    E(\bm{\Theta}) = \frac{1}{N_{s}N_{m}} \| \mathbf{X} - \mathbf{X}^{\ast} \|_{F}^{2}
    \label{equation:loss_function_matrix}
\end{equation}
where $\mathbf{X}^{\ast}$ is the ground-truth multimodal graph signal.
Note that all components are easily (sub-)differentiable and bounded-input-bounded-output, and parameters can be trained using a general optimizer such as Adam \cite{DBLP:journals/corr/KingmaB14}.
Therefore, all the parameters can be trained by backpropagation.

{In the original model-based optimization problem, ten parameters are tuned: Six in \eqref{equation:objective_proposed} ($\alpha_s, \beta_s, \gamma_s, \alpha_m, \beta_m, \gamma_m$), three in the PDS algorithm (number of iterations, step size, and tolerance), and one for the number of iterations $T$. 
On the other hand, in the unrolled counterpart, the six parameters ($\alpha_s, \dots, \gamma_m$) are trained, so the number of parameters to be manually tuned is reduced to five: three for the internal parameters of the PDS algorithm's internal parameters, one for the number of layers $T$, and one in the learning rate of the optimizer during training.
In fact, these five parameters to be tuned in our method are less sensitive than the original regularization strengths. 

\begin{algorithm}[tb]
    \caption{PDS algorithm for solving \eqref{equation:objective_function_sub_vech_indicator}}
    \label{algorithm:pds_form_e}
    \textbf{Input: } $
    \mathbf{X},
    \bm{\Phi},
    \bm{\Psi},
    \alpha=\alpha^{(t)}, \beta=\beta^{(t)}, \gamma=\gamma^{(t)}, \theta, \textrm{tolerance }\epsilon
    $
    \begin{algorithmic}[1]
        \State \textbf{Initialization: } $\bm{\ell}^{(0)}= -\mathbf{1}/10, \mathbf{b}_{0}^{(0)} = \bm{\Psi} \bm{\ell}^{(0)}$
        \For{$i = 0, \ldots, i_{\textrm{max}}-1$}
            \State $\mathbf{z}^{(i)} \gets \bm{\ell}^{(i)} - \theta ( \alpha \bm{\Phi}^{\top} \textrm{vec}(\mathbf{XX}^{\top}) + 2 \gamma \bm{\ell}^{(i)} + \bm{\Psi}^{\top} \mathbf{b}_{0}^{(i)})$
            \State $\mathbf{z}_{0}^{(i)} \gets \mathbf{b}_{0}^{(i)} + \theta \bm{\Psi} \bm{\ell}^{(i)}$
            \State $\mathbf{p}^{(i)} \gets \textrm{prox}_{\theta \iota_{\cdot \leq \mathbf{0}}} (\mathbf{z}^{(i)})$
            \State $\mathbf{p}_{0}^{(i)} \gets \mathbf{z}_{0}^{(i)} - \theta \textrm{prox}_{\frac{1}{\theta} ( - \beta \mathbf{1}_{\mathcal{V}}^{\top} \log(\cdot))} \left (\frac{\mathbf{z}_{0}^{(i)}}{\theta} \right)$
            \State $\mathbf{q}^{(i)} \gets \mathbf{p}^{(i)} - \theta ( \alpha \bm{\Phi}^{\top} \textrm{vec}(\mathbf{XX}^{\top}) + 2 \gamma \mathbf{p}^{(i)} + \bm{\Psi}^{\top} \mathbf{p}_{0}^{(i)}) $
            \State $\mathbf{q}_{0}^{(i)} \gets \mathbf{p}_{0}^{(i)} + \theta \bm{\Psi} \mathbf{p}^{(i)}$
            \State $\bm{\ell}^{(i+1)} \gets \bm{\ell}^{(i)} - \mathbf{z}^{(i)} + \mathbf{q}^{(i)}$
            \State $\mathbf{b}_{0}^{(i+1)} \gets \mathbf{b}_{0}^{(i)} - \mathbf{z}_{0}^{(i)} + \mathbf{q}_{0}^{(i)}$
            \If{$\| \bm{\ell}^{(i+1)} - \bm{\ell}^{(i)} \| / \| \bm{\ell}^{(i)} \| < \epsilon $}
                \State \textbf{break}
            \EndIf
        \EndFor
    \end{algorithmic}
    \textbf{Output: } $\bm{\ell}^{(i+1)}$
\end{algorithm}

\begin{algorithm}[tb]
    \caption{MGSD-LLap-DAU: Multimodal graph signal denoising with simultaneously learning Laplacian matrix using deep algorithm unrolling}
    \label{algorithm:MGSD-LLap-DAU}
    \textbf{Input: } $\mathbf{X}_{s}^{(0)}=\mathbf{Y}, T, \{ \alpha_{e}^{(t)}, \beta_{e}^{(t)}, \gamma_{e}^{(t)} \}_{t=0}^{T-1}~(e \in \{m,s\})$
    \begin{algorithmic}[1]
        \ForAll{$t=0,\ldots,T-1$}
            \State $\mathbf{L}_{m}^{(t+1)} \gets \argmin_{L_{m}} f_{m}((\mathbf{X}_{s}^{(t)})^{\top}, \mathbf{L}_{m}) $
            \State $\mathbf{X}_{m}^{(t+1)} \gets (\mathbf{I}_{N_{m}} + \alpha_{m}^{(t)} \mathbf{L}_{m}^{(t+1)})^{-1}\mathbf{Y}^{\top}$
            \State $\mathbf{L}_{s}^{(t+1)} \gets \argmin_{L_{s}} f_{s}((\mathbf{X}_{m}^{(t+1)})^{\top}, \mathbf{L}_{s})  $
            \State $\mathbf{X}_{s}^{(t+1)} \gets (\mathbf{I}_{N_{s}} + \alpha_{s}^{(t)} \mathbf{L}_{s}^{(t+1)})^{-1} (\mathbf{X}_{m}^{(t+1)})^{\top}$
        \EndFor
    \end{algorithmic}
    \textbf{Output: } $\mathbf{X}=\mathbf{X}_{s}^{(T)}, \mathbf{L}_{m}^{(T)}, \mathbf{L}_{s}^{(T)}$
\end{algorithm}

\begin{figure}[t]
    \centering
    \includegraphics[width=0.95\hsize]{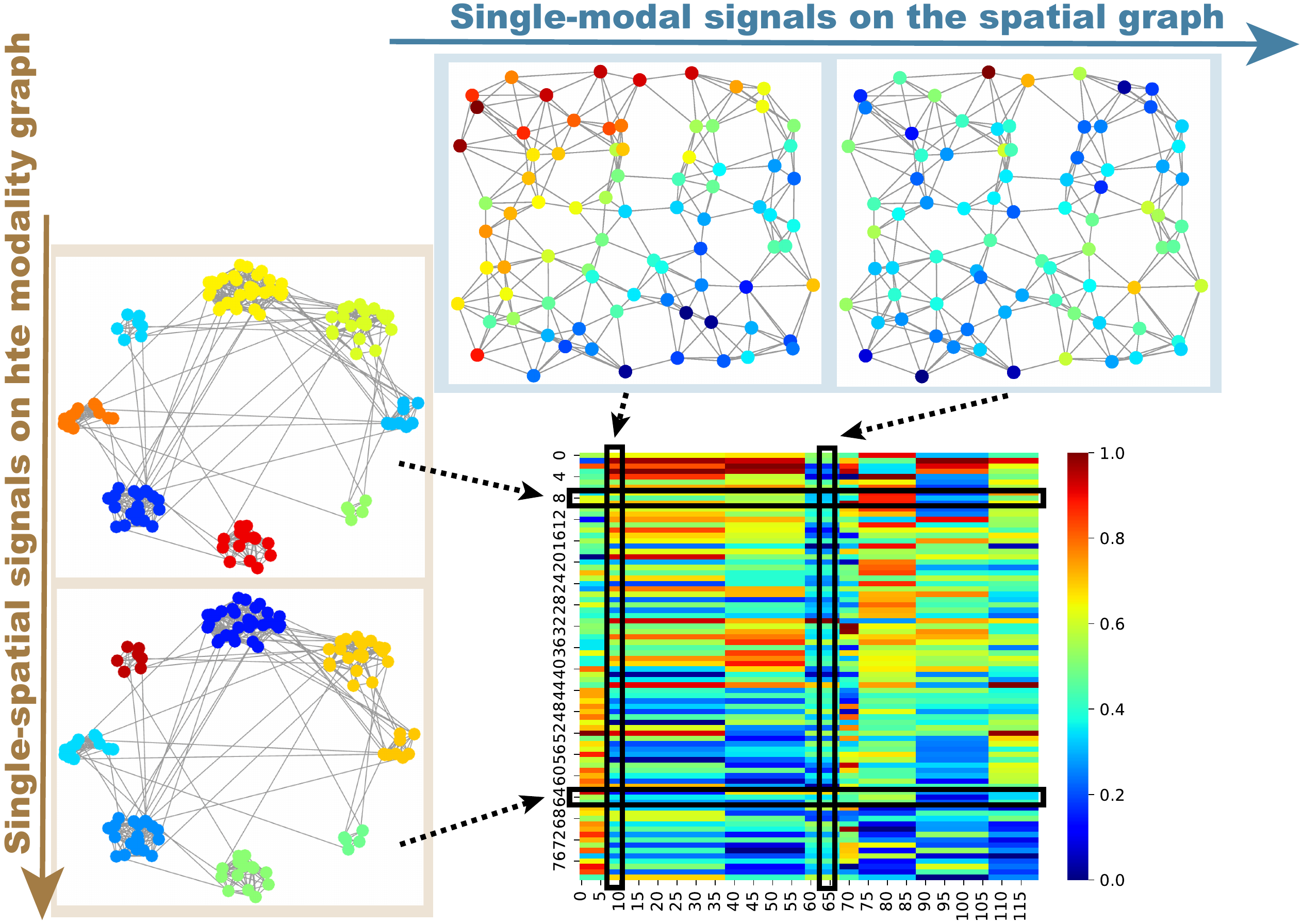}
    \caption{Visualization of synthetic multimodal graph signals.}
    \label{figure:visualization_multimodal_graph_signal}
\end{figure}

\begin{table}[t]
\centering
\caption{RMSE of denoised signals for synthetic dataset.}
\label{table:rmse_Synthetic}
\begin{tabular}{l|ccccc}
\bhline{1.1pt}
Methods 	\textbackslash $\sigma$ & 0.10 & 0.15 & 0.20 & 0.25 & 0.30 \\ \hline\hline
GLPF & 0.084 & 0.122 & 0.171 & 0.224 & 0.283 \\
HD & 0.203 & 0.198 & 0.188 & 0.183 & 0.194 \\
SVDS & 0.100 & 0.149 & 0.199 & 0.249 & 0.299 \\ \hline
AE & 0.261 & 0.261 & 0.261 & 0.261 & 0.261 \\
GCN & 0.209 & 0.205 & 0.204 & 0.200 & 0.195 \\ \hline
TGSR-DAU & 0.068 & 0.091 & 0.117 & 0.138 & 0.155 \\
MGSD-LLAP-DAU & \textbf{0.030} & \textbf{0.039} & \textbf{0.051} & \textbf{0.060} & \textbf{0.076} \\ \bhline{1.1pt}
\end{tabular}
\end{table}

\begin{figure*}[t]
    \begin{minipage}[t]{0.30\linewidth}
        \centering
        \includegraphics[width=5cm]{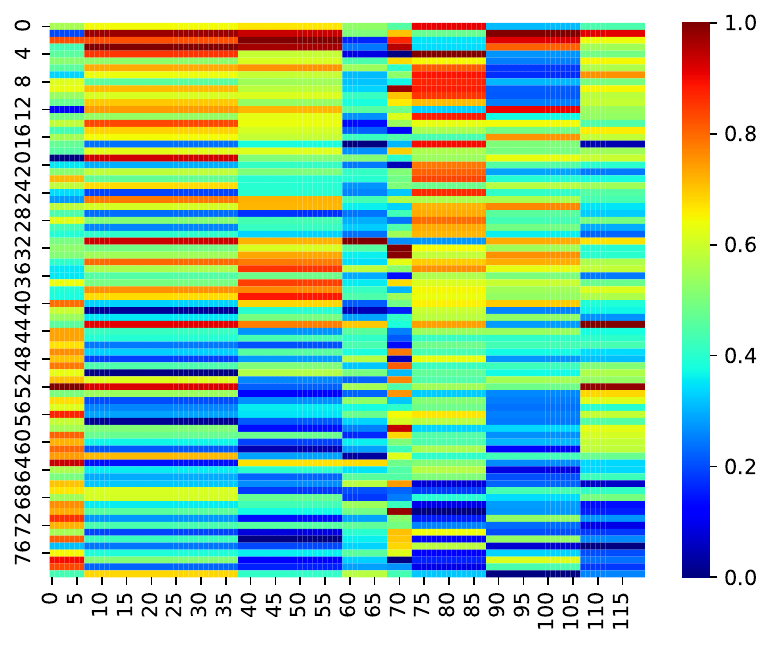}
        \subcaption{Original}
    \end{minipage}%
    \hfill
    \begin{minipage}[t]{0.30\linewidth}
        \centering
        \includegraphics[width=5cm]{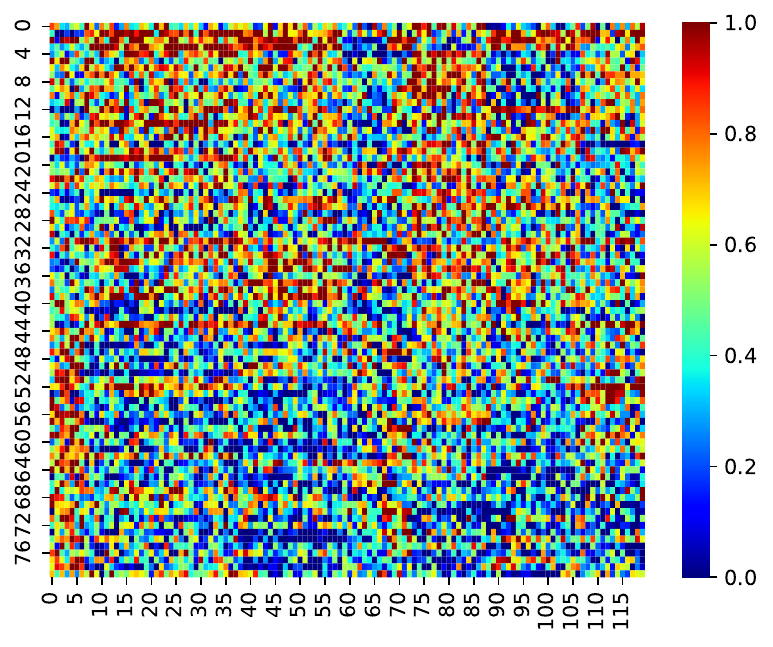}
        \subcaption{Observed (0.30)}
    \end{minipage}
    \hfill
    \begin{minipage}[t]{0.30\linewidth}
        \centering
        \includegraphics[width=5cm]{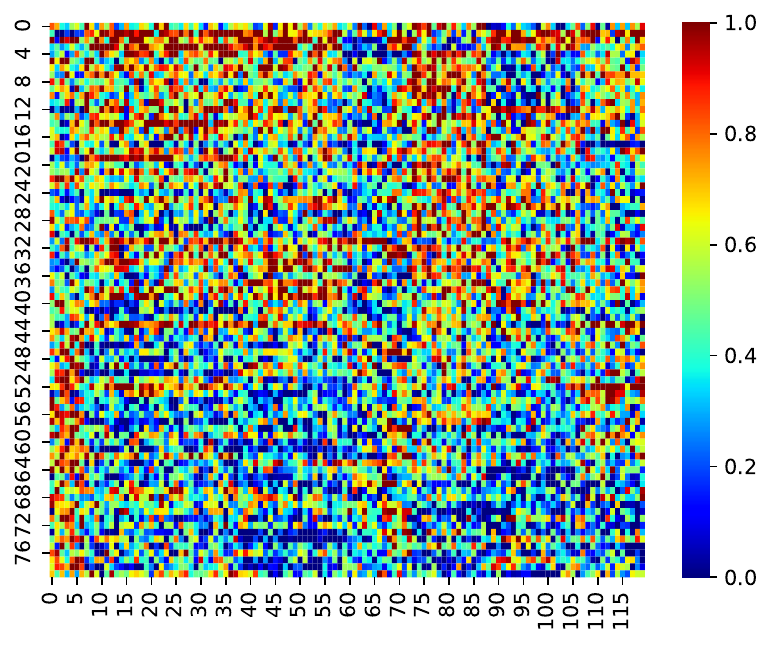}
        \subcaption{GLPF (0.27)}
    \end{minipage}
    \hfill
    \begin{minipage}[t]{0.30\linewidth}
        \centering
        \includegraphics[width=5cm]{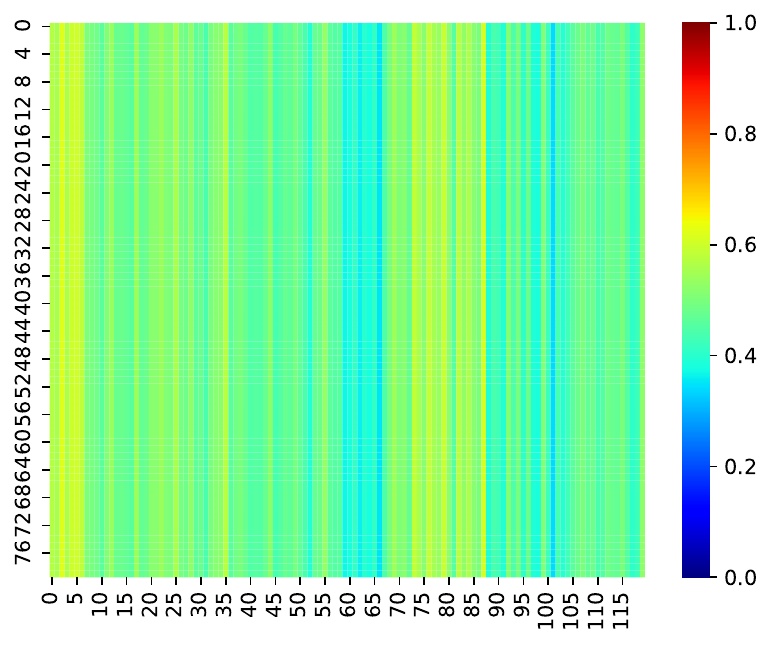}
        \subcaption{HD (0.22)}
    \end{minipage}
    \hfill
    \begin{minipage}[t]{0.30\linewidth}
        \centering
        \includegraphics[width=5cm]{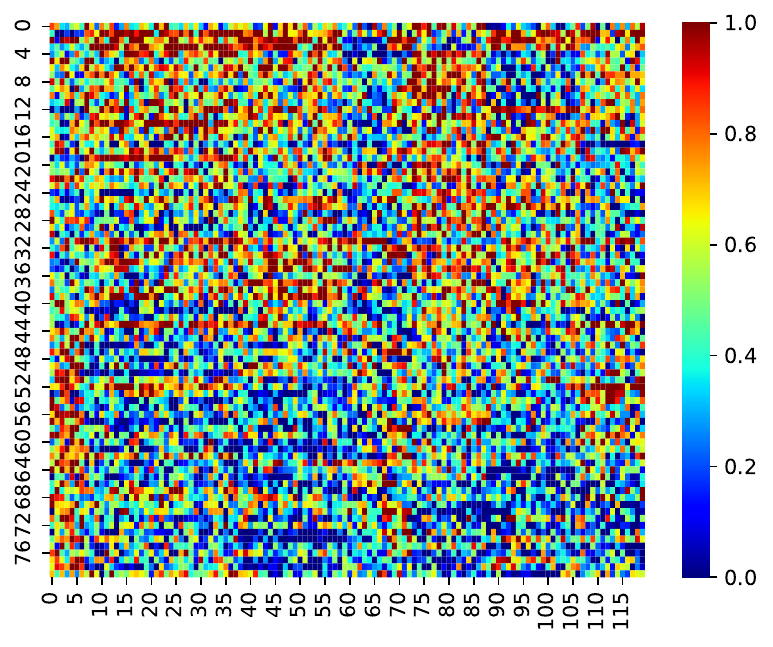}
        \subcaption{SVDS (0.30)}
    \end{minipage}
    \hfill
    \begin{minipage}[t]{0.30\linewidth}
        \centering
        \includegraphics[width=5cm]{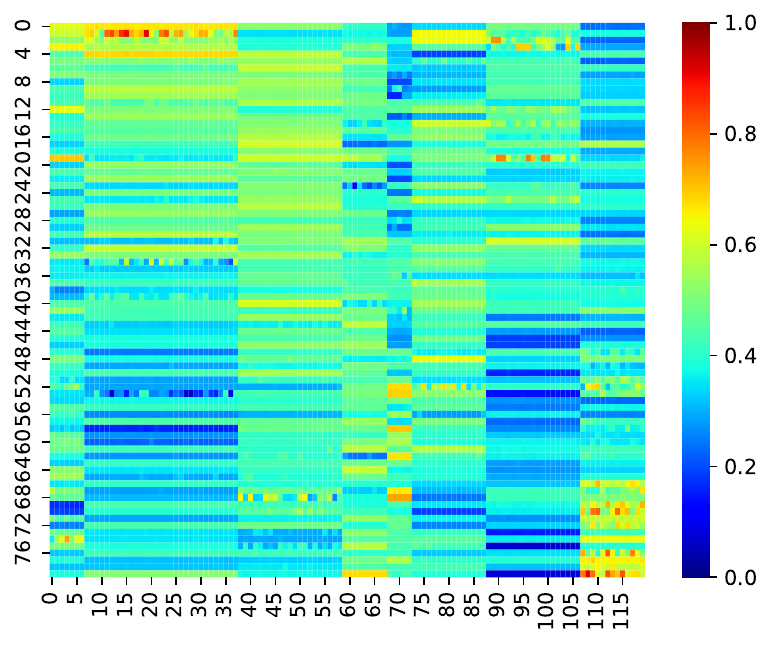}
        \subcaption{AE (0.24)}
    \end{minipage}
    \hfill
    \begin{minipage}[t]{0.30\linewidth}
        \centering
        \includegraphics[width=5cm]{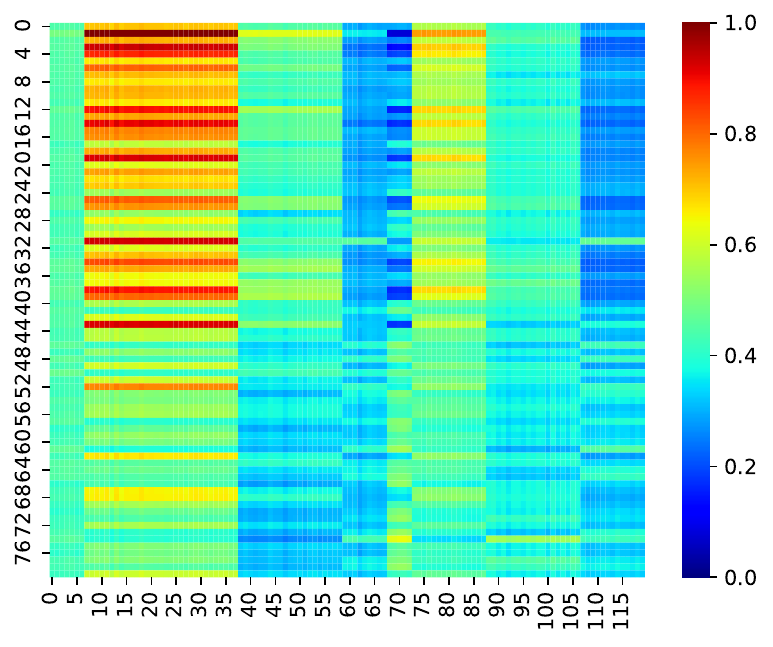}
        \subcaption{GCN (0.19)}
    \end{minipage}
    \hfill
    \begin{minipage}[t]{0.30\linewidth}
        \centering
        \includegraphics[width=5cm]{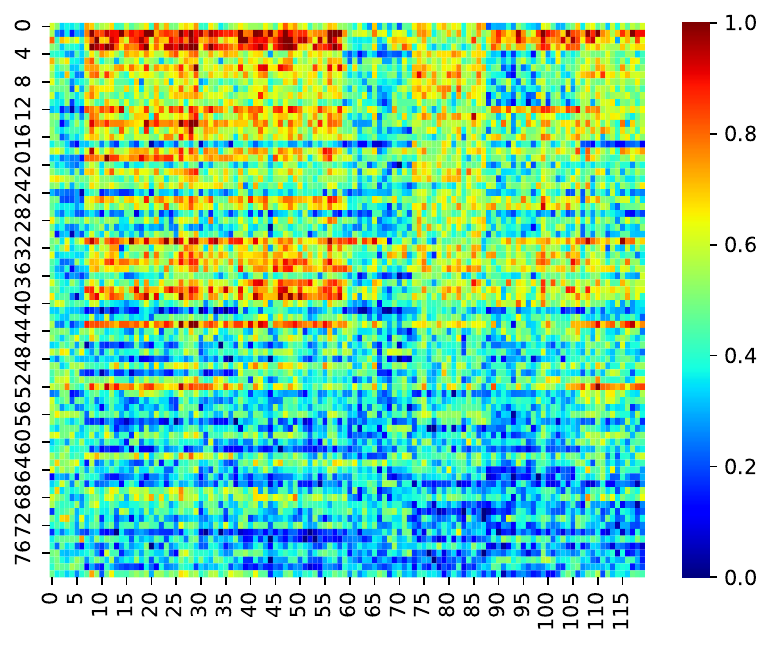}
        \subcaption{TGSR-DAU (0.16)}
    \end{minipage}
    \hfill
    \begin{minipage}[t]{0.30\linewidth}
        \centering
        \includegraphics[width=5cm]{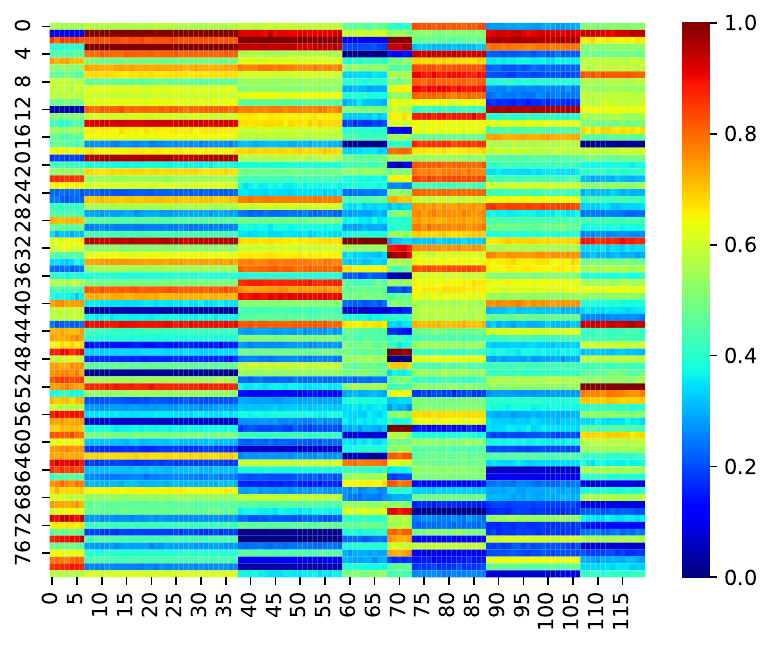}
        \subcaption{ MGSD-LLAP-DAU  (0.07)}
    \end{minipage}
    \caption{Denoising results for synthetic dataset ($\sigma=0.30$). The vertical indices correspond to the spatial graph numbers and the horizontal ones are the modality. 
    Parenthesized values in subcaptions are RMSE.
    }
    \label{fig:visualization_syn}
\end{figure*}

\begin{figure*}[t]
    \begin{minipage}[t]{0.2\linewidth}
        \centering
        \includegraphics[width=3.5cm]{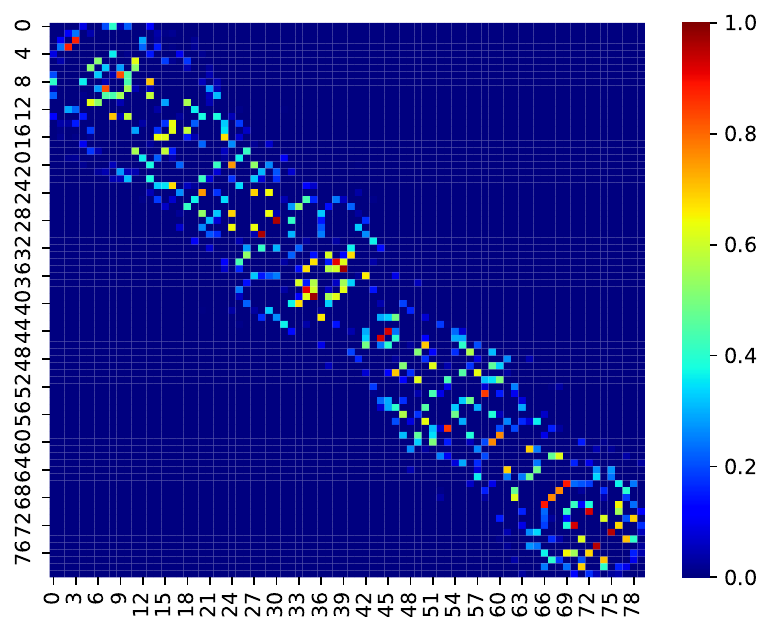}
        \subcaption{Ground-truth}
    \end{minipage}%
    \hfill
    \begin{minipage}[t]{0.2\linewidth}
        \centering
        \includegraphics[width=3.5cm]{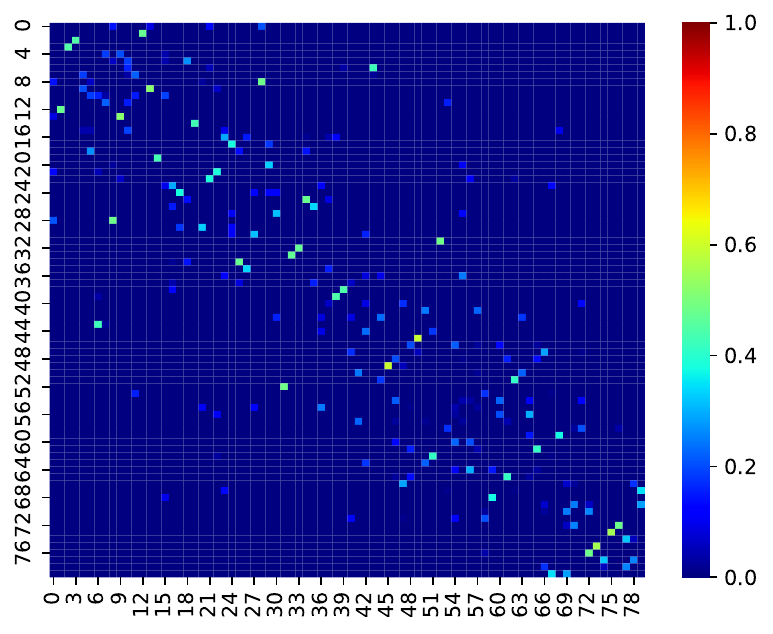}
        \subcaption{Layer 1}
    \end{minipage}%
    \hfill
    \begin{minipage}[t]{0.2\linewidth}
        \centering
        \includegraphics[width=3.5cm]{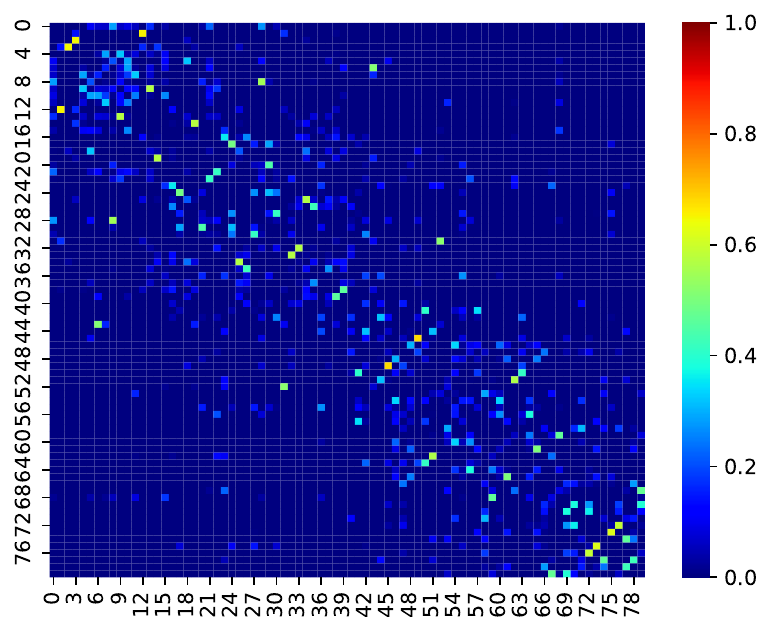}
        \subcaption{Layer 2}
    \end{minipage}%
    \hfill
    \begin{minipage}[t]{0.2\linewidth}
        \centering
        \includegraphics[width=3.5cm]{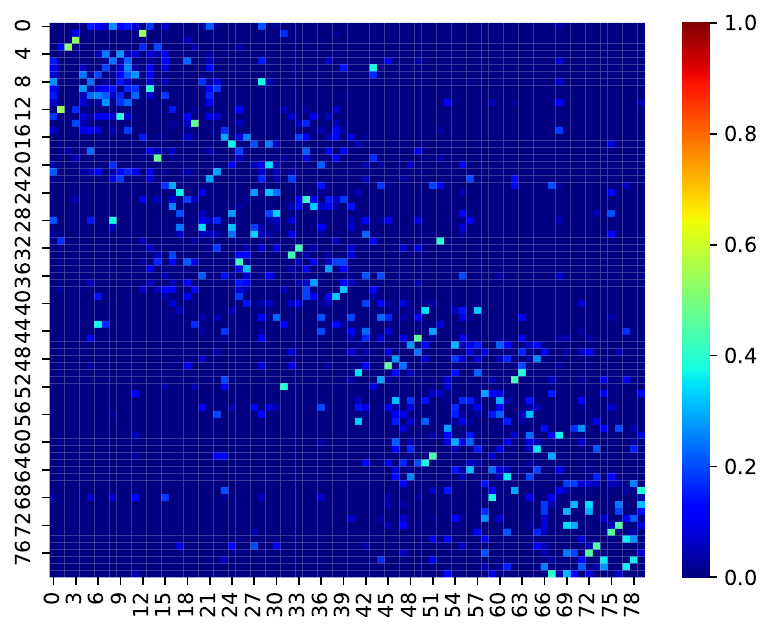}
        \subcaption{Layer 3}
    \end{minipage}%
    \hfill
    \begin{minipage}[t]{0.2\linewidth}
        \centering
        \includegraphics[width=3.5cm]{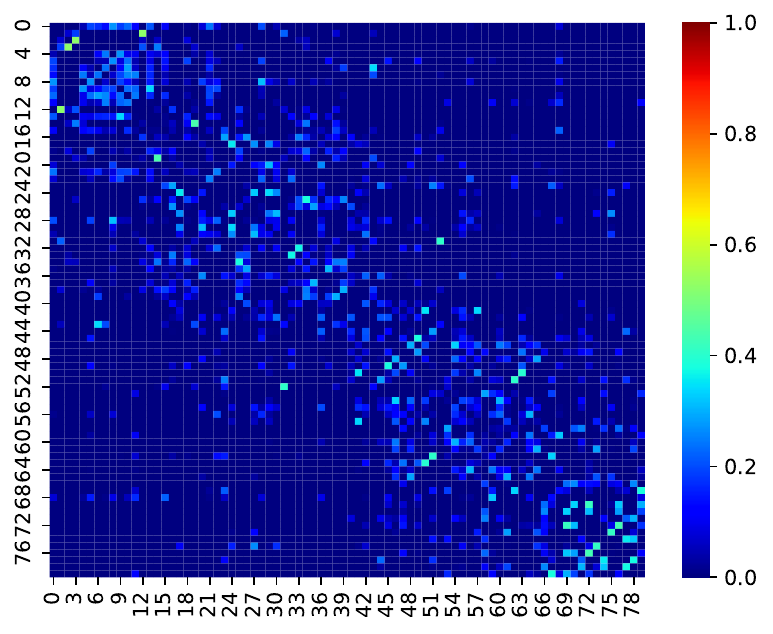}
        \subcaption{Layer 4}
    \end{minipage}%
    \hfill
    \begin{minipage}[t]{0.2\linewidth}
        \centering
        \includegraphics[width=3.5cm]{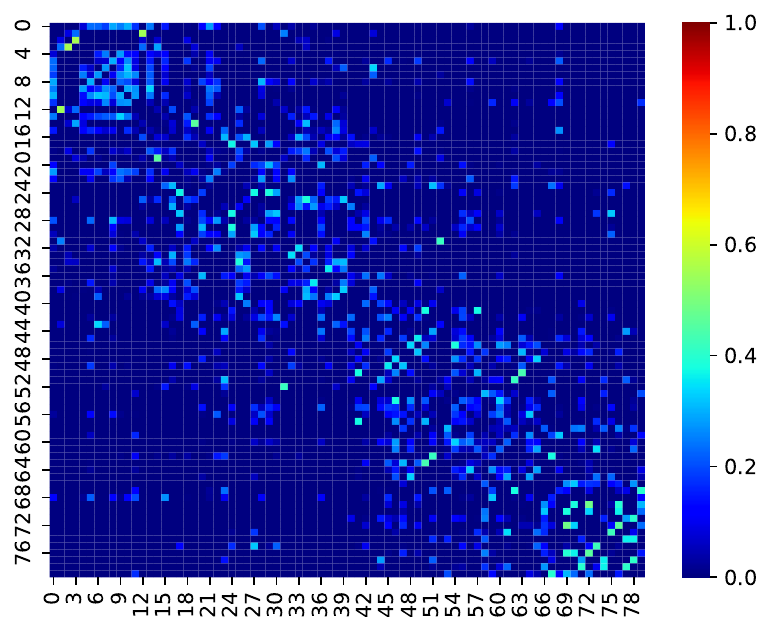}
        \subcaption{Layer 5}
    \end{minipage}%
    \hfill
    \begin{minipage}[t]{0.2\linewidth}
        \centering
        \includegraphics[width=3.5cm]{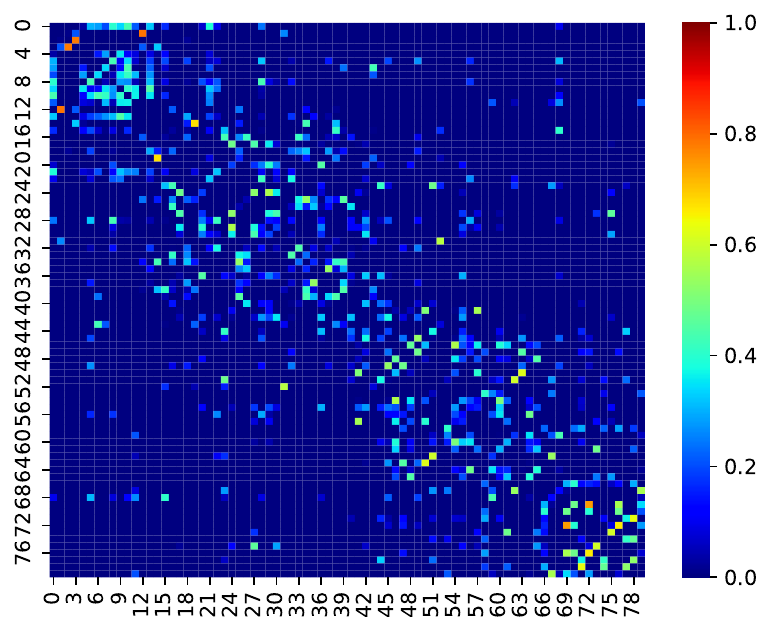}
        \subcaption{Layer 6}
    \end{minipage}%
    \hfill
    \begin{minipage}[t]{0.2\linewidth}
        \centering
        \includegraphics[width=3.5cm]{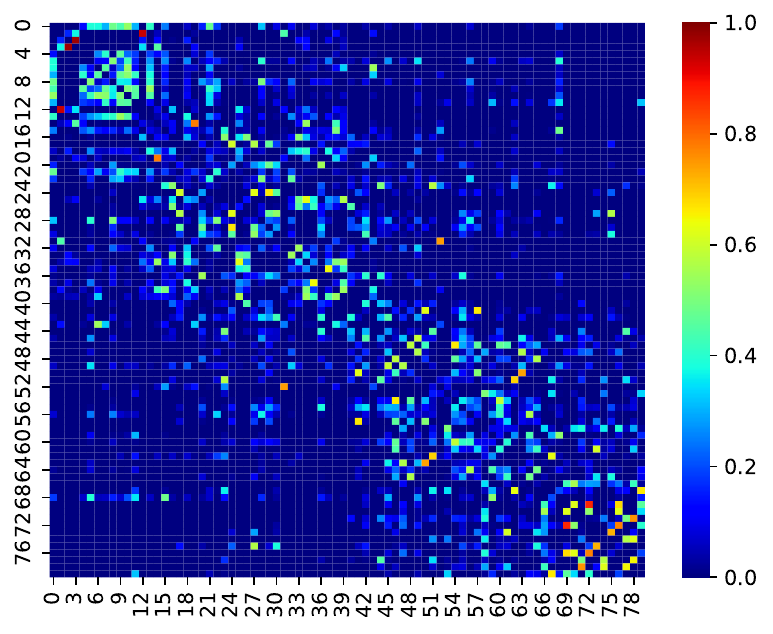}
        \subcaption{Layer 7}
    \end{minipage}%
    \hfill
    \begin{minipage}[t]{0.2\linewidth}
        \centering
        \includegraphics[width=3.5cm]{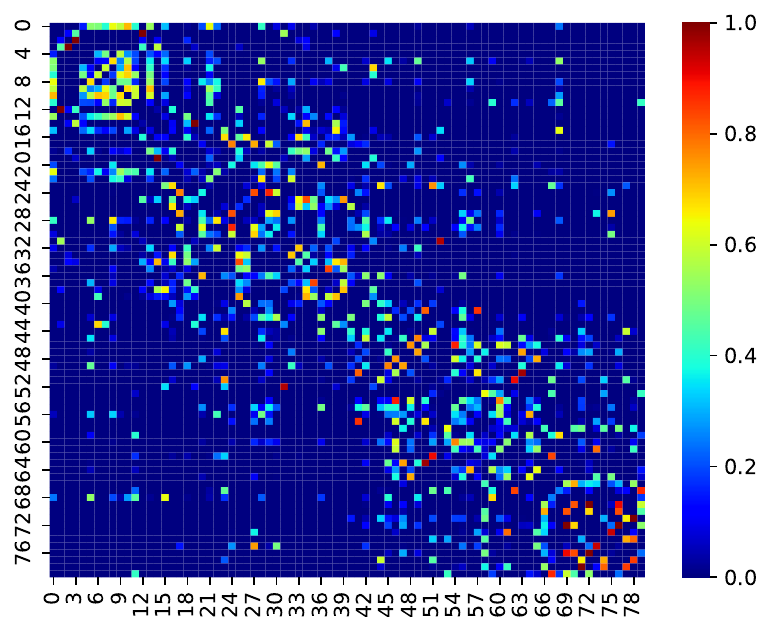}
        \subcaption{Layer 8}
    \end{minipage}%
    \hfill
    \begin{minipage}[t]{0.2\linewidth}
        \centering
        \includegraphics[width=3.5cm]{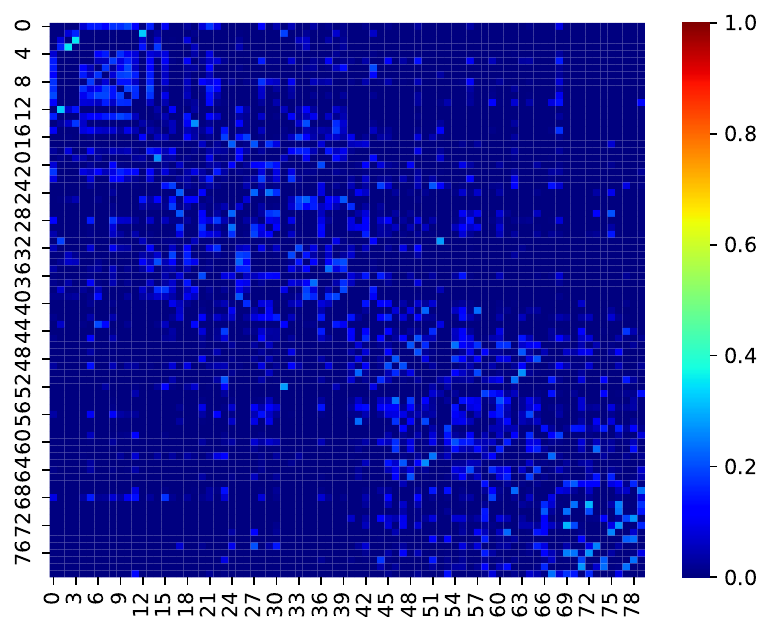}
        \subcaption{Layer 9}
    \end{minipage}%
    \hfill
    \caption{Visualization of learned spatial graph $\mathbf{W}_{s}$ for synthetic dataset. 
    }
    \label{fig:Ws_syn}
\end{figure*}

\begin{figure*}[t]
    \begin{minipage}[t]{0.2\linewidth}
        \centering
        \includegraphics[width=3.5cm]{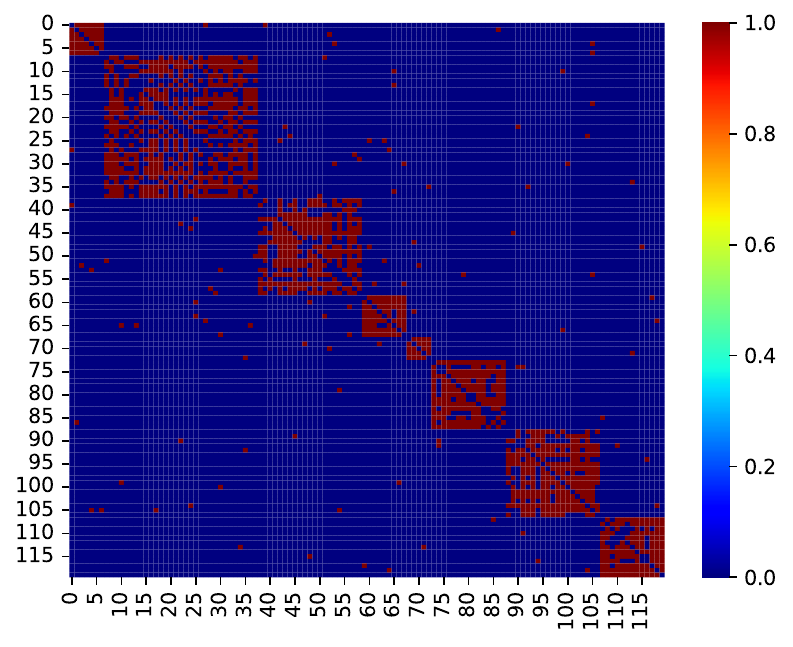}
        \subcaption{Ground-truth}
    \end{minipage}%
    \hfill
    \begin{minipage}[t]{0.2\linewidth}
        \centering
        \includegraphics[width=3.5cm]{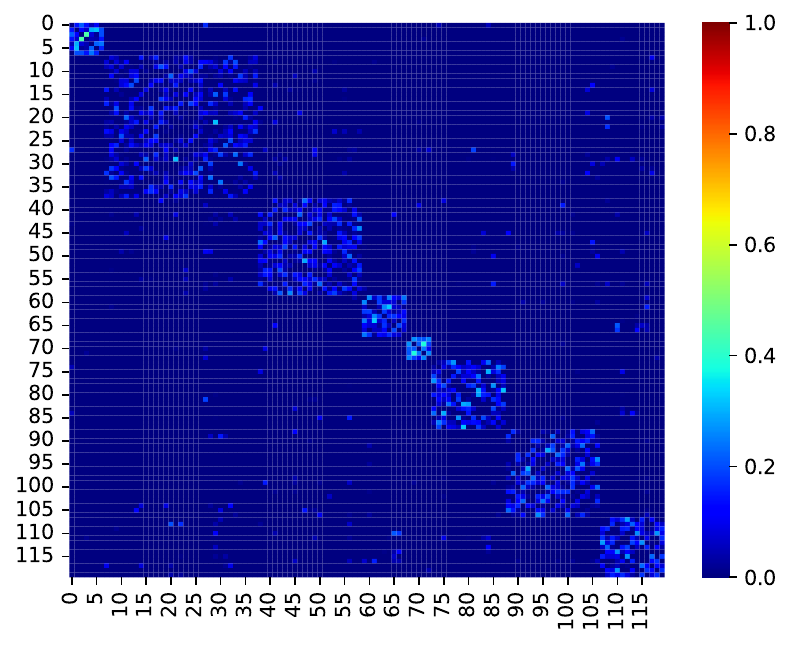}
        \subcaption{Layer 1}
    \end{minipage}%
    \hfill
    \begin{minipage}[t]{0.2\linewidth}
        \centering
        \includegraphics[width=3.5cm]{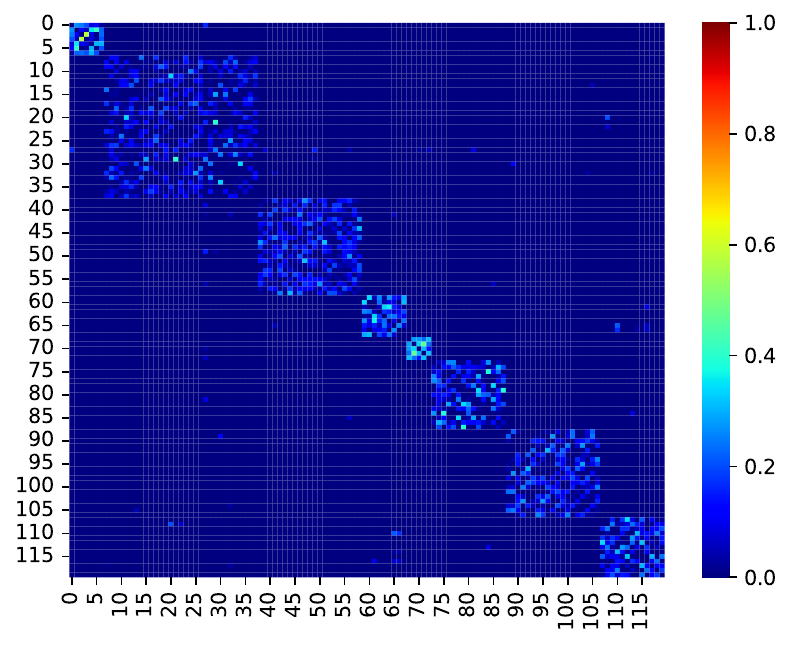}
        \subcaption{Layer 2}
    \end{minipage}%
    \hfill
    \begin{minipage}[t]{0.2\linewidth}
        \centering
        \includegraphics[width=3.5cm]{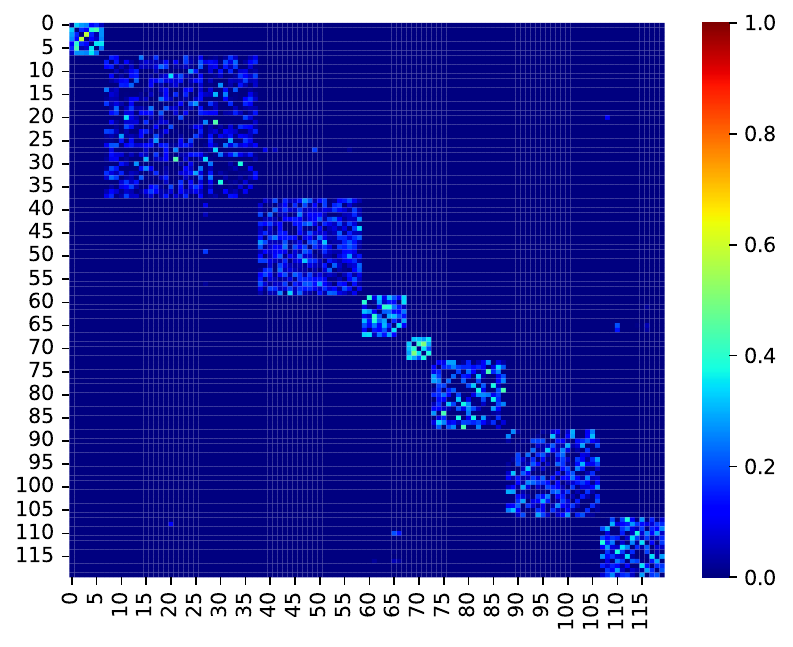}
        \subcaption{Layer 3}
    \end{minipage}%
    \hfill
    \begin{minipage}[t]{0.2\linewidth}
        \centering
        \includegraphics[width=3.5cm]{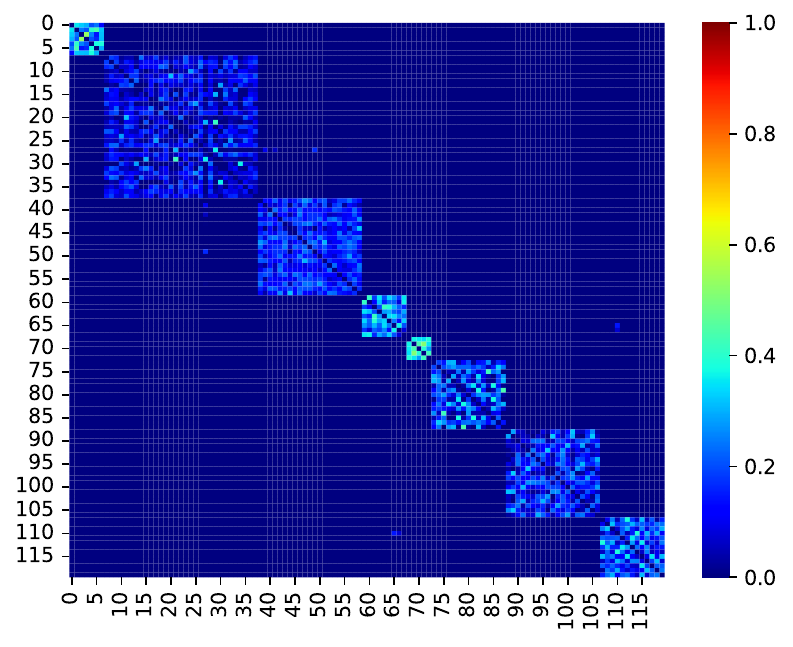}
        \subcaption{Layer 4}
    \end{minipage}%
    \hfill
    \begin{minipage}[t]{0.2\linewidth}
        \centering
        \includegraphics[width=3.5cm]{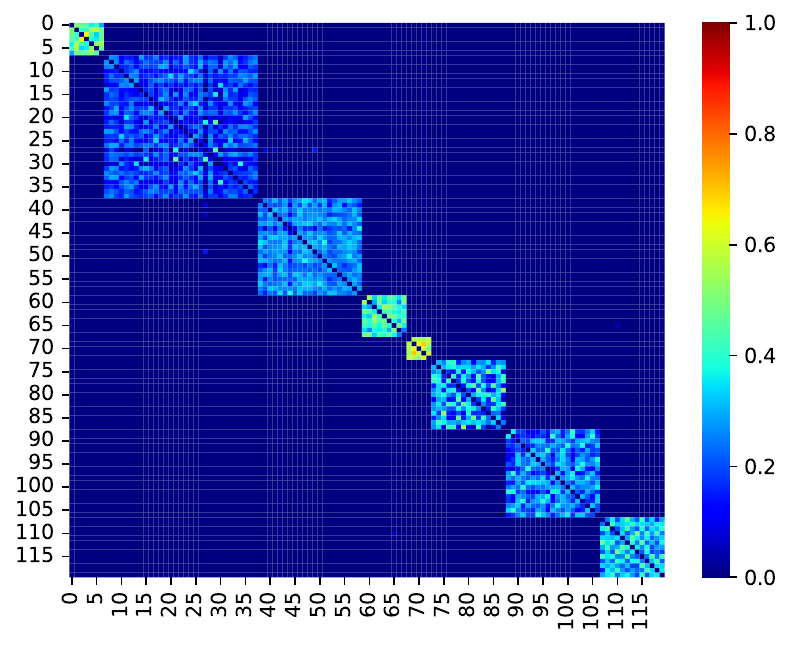}
        \subcaption{Layer 5}
    \end{minipage}%
    \hfill
    \begin{minipage}[t]{0.2\linewidth}
        \centering
        \includegraphics[width=3.5cm]{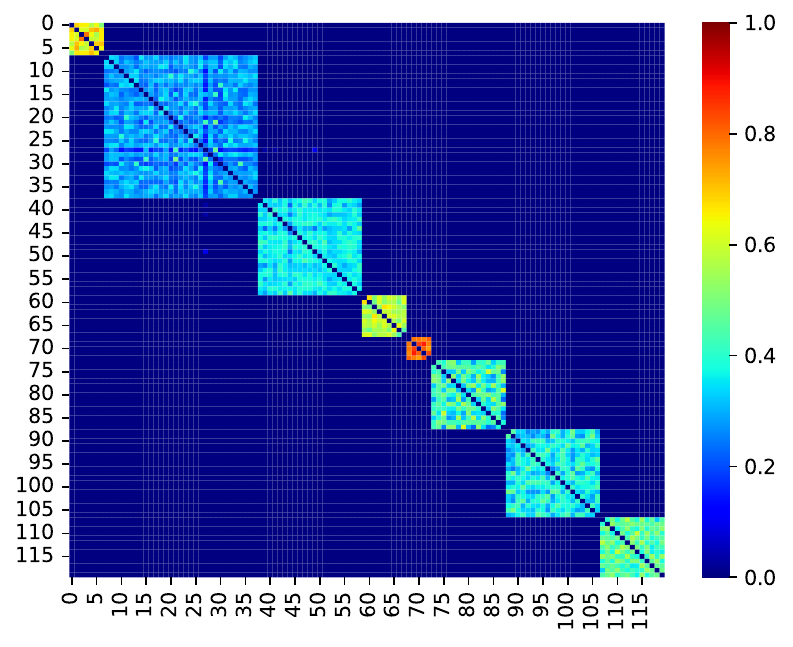}
        \subcaption{Layer 6}
    \end{minipage}%
    \hfill
    \begin{minipage}[t]{0.2\linewidth}
        \centering
        \includegraphics[width=3.5cm]{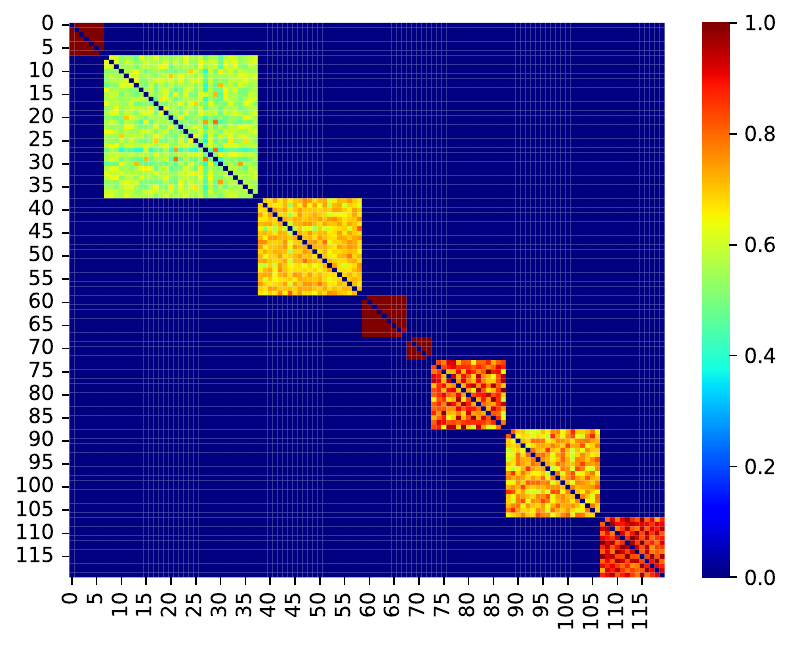}
        \subcaption{Layer 7}
    \end{minipage}%
    \hfill
    \begin{minipage}[t]{0.2\linewidth}
        \centering
        \includegraphics[width=3.5cm]{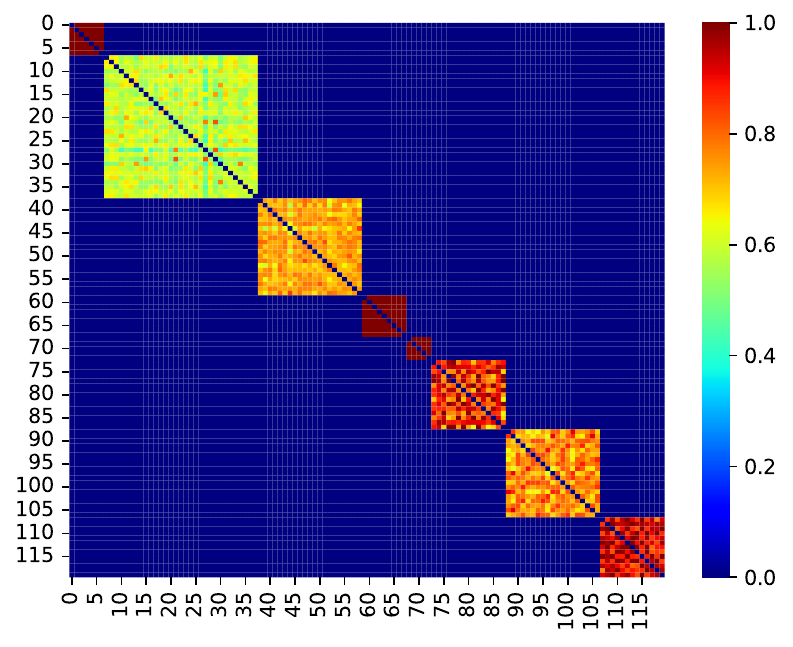}
        \subcaption{Layer 8}
    \end{minipage}%
    \hfill
    \begin{minipage}[t]{0.2\linewidth}
        \centering
        \includegraphics[width=3.5cm]{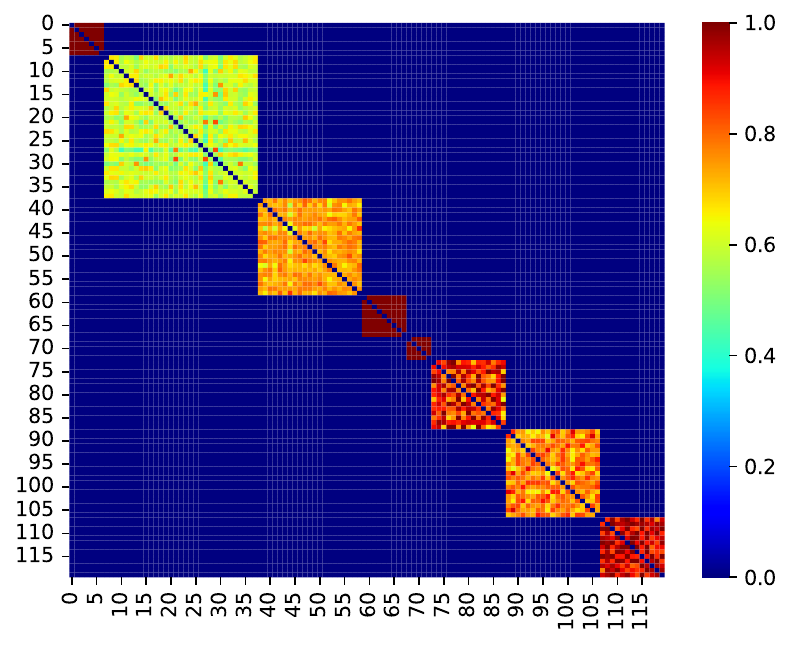}
        \subcaption{Layer 9}
    \end{minipage}%
    \hfill
    \caption{Visualization of learned modality graph $\mathbf{W}_{m}$ for synthetic dataset. 
    }
    \label{fig:Wm_syn}
\end{figure*}

\section{DENOISING EXPERIMENTS} \label{section:exp}
In this section, we perform denoising experiments of multimodal signals with synthetic and real-world datasets\footnote{Code for the experiments is available at \url{https://github.com/kojima-msp/MGSD_LLAP_DAU}.}.

\subsection{SETUP}

\subsubsection{DATASETS}
We use the following two datasets.

\noindent
\textbf{Synthetic Dataset:}
As a ground-truth twofold graph, we use a random sensor graph as a spatial graph $\mathcal{G}_{s}$ and a community graph \cite{perraudinGSPBOXToolboxSignal2016}.
This setup reflects a typical measurement condition of wireless sensor networks or IoT with environmental sensors.

The spatial graph $\mathcal{G}_{s}$ is constructed as a $k$-nearest neighbor (kNN) graph with $k=6$, based on the coordinates of the nodes.
The edge weights of $\mathcal{G}_{s}$ are determined based on their Euclidean distances in space and are rescaled to $[0,1]$.

We generated a modality graph $\mathcal{G}_{m}$ with eight clusters ($P=8$).
First, all nodes within the same cluster are connected.
Then, all pairs of inter-cluster nodes are connected with a probability of $1/N_m$.
Edge weights for the modality graph are unweighted, i.e., $w_{i,j} = 1$. 
We set $N_{s}=80$ and $N_{m}=120$.

The edge weights of $\mathcal{G}_{m}$ are set to be unweighted ($\{0, 1\}$) based on the cluster.

Synthetic multimodal graph signals are obtained through two steps:
\begin{enumerate}
    \item A spatial graph signal $\tilde{\mathbf{x}}_{p} \in \mathbb{R}^N$ is designed such that it varies smoothly on $\mathcal{G}_{s}$.
    It has been made to conform to a Gaussian Markov Random Field (GMRF) process \cite{rueGaussianMarkovRandom2005}, i.e., $\tilde{\mathbf{x}}_{p} \sim \mathcal{N}(\mathbf{0}, \mathbf{L}_{s}^{\dagger})$, where $\mathbf{L}_{s}^{\dagger}$ is the Moore-Penrose pseudoinverse of $\mathbf{L}_{s}$.
    \item Based on $\tilde{\mathbf{x}}_{p}$, we then modify the signal values to be correlated along the modality direction.
    We assume that multimodal signals are piecewise constant \cite{chenRepresentationsPiecewiseSmooth2016} on the modality graph. 
    Therefore, the measurement matrix $\mathbf{X}$ is given by $\mathbf{X} = \begin{bmatrix} \tilde{\mathbf{x}}_{0} \mathbf{1}_{\mathcal{C}_{0}}^{\top} \cdots \tilde{\mathbf{x}}_{P-1} \mathbf{1}_{\mathcal{C}_{P-1}}^{\top}\end{bmatrix}$ where $\mathcal{C}_{p}$ is the set of nodes in the $p$th cluster of $\mathcal{G}_{m}$.
\end{enumerate}
We visualize synthetic multimodal graph signals in Fig.~\ref{figure:visualization_multimodal_graph_signal}.

For the experiments, we generate ten multimodal graphs.
For each graph, experiments with five noise levels, $\sigma \in \{0.10, 0.15, 0.20, 0.25, 0.30\}$, are performed.
Parameters are determined with a $2\times 2$ cross-validation.

\noindent
\textbf{Real-world Dataset:}
We also perform denoising experiments using Japanese climate data publicly available from the Japan Meteorological Agency (JMA\footnote{JMA HP: \url{https://www.data.jma.go.jp/obd/stats/etrn/index.php}}).
In this experiment, we consider modality as seasons. 

We utilize the temperature data consisting of daily average temperatures spanning ten years from 2013 to 2022, measured at 62 locations across Japan. 
To align the number of days in each year, data for February 29th in leap years have been excluded. 
The 62 locations were selected from observatories in Japan that had non-missing temperature data for the entire ten-year period. 

The observed signals are made by the following two steps:
\begin{enumerate}
    \item By performing five-day averaging, we have 73 sets of temperature data. This forms the original signals $\mathbf{X} \in \mathbb{R}^{62 \times 73}$.
    \item We then add AWGN ($\sigma \in \{3,5,7,9\}$) to the data to obtain the observed signal $\mathbf{Y}$. 
\end{enumerate}
Finally, we repeat the above two steps ten times to arrange $40$ pairs of $\{\mathbf{X}, \mathbf{Y}\}$. 
We performed $2\times 2$ cross-validation for the dataset. 

\begin{figure}[tbp]
    \centering
    \includegraphics[width=0.7\linewidth]{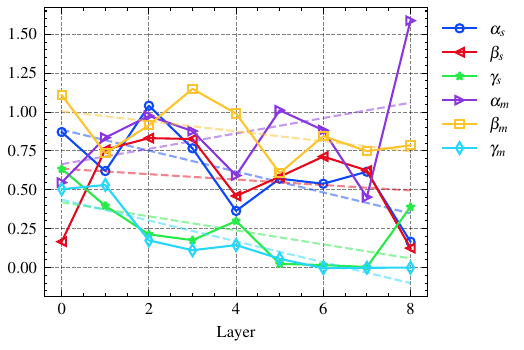}
    \caption{Values of the learned parameters for each layer $t$. The model was trained on the synthetic dataset. The dashed lines indicate the results of linear regression lines for the corresponding each parameters.}
    \label{fig:learned_parameters}
\end{figure}

\begin{table}[t]
\centering
\caption{RMSE of denoised signals for synthetic dataset.}
\label{table:ablation_rmse}
\begin{tabular}{l|ccccc}
\bhline{1.1pt}
Methods 	\textbackslash $\sigma$ & 0.10 & 0.15 & 0.20 & 0.25 & 0.30 \\ \hline\hline
Oracle   & 0.056 & 0.064 & 0.073 & 0.083 & 0.093 \\ \hline
Learned ($\mathbf{L}_s$ and $\mathbf{L}_m$) & 0.117 & 0.126 & 0.140 & 0.156 & 0.174 \\
Learned ($\mathbf{L}_s$) & 0.076 & 0.085 & 0.099 & 0.115 & 0.136 \\
Learned ($\mathbf{L}_m$)  & 0.119 & 0.128 & 0.141 & 0.156 & 0.172 \\ 
Proposed & \textbf{0.030} & \textbf{0.039} & \textbf{0.051} & \textbf{0.060} & \textbf{0.076} \\
\bhline{1.1pt}
\end{tabular}
\end{table}

\begin{figure}
    \centering
    \includegraphics[width=0.7\linewidth]{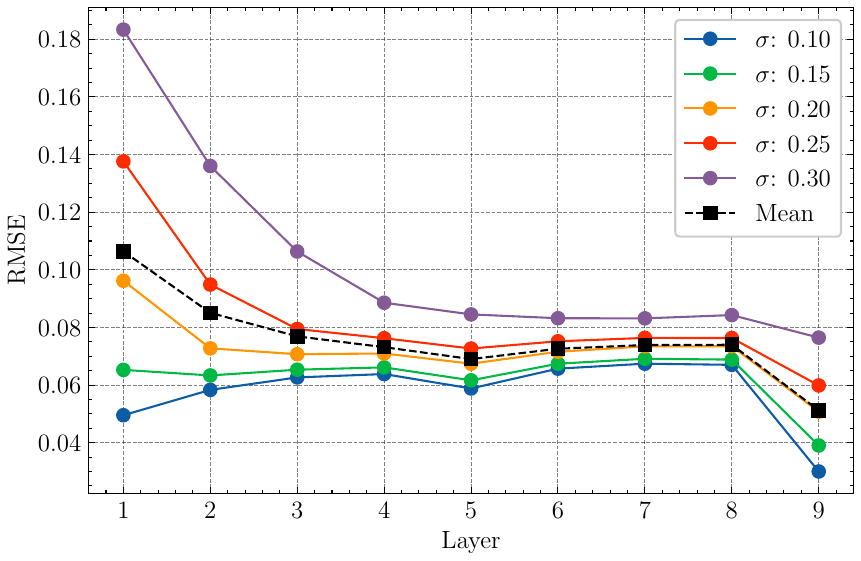}
    \caption{Layer-wise root mean squared error (RMSE) for different noise levels on the synthetic dataset. }
    \label{fig:rmse_vs_layer}
\end{figure}

\begin{table}[t]
\centering
\caption{RMSE of denoised signals for real-world dataset.}
\label{table:rmse_Weather}
\begin{tabular}{l|cccc}
\bhline{1.1pt}
Methods \textbackslash $\sigma$ & 3 & 5 & 7 & 9 \\
\hline \hline
GLPF & 1.937 & 3.154 & 5.346 & 7.925 \\
HD & 2.982 & 3.800 & 4.838 & 6.046 \\
SVDS & 2.988 & 4.990 & 6.952 & 9.006 \\ \hline
AE & 3.776 & 3.776 & 3.776 & 3.776 \\
GCN & 3.437 & 3.266 & 3.406 & 4.082 \\ \hline
TGSR-DAU & 2.394 & 2.834 & 3.386 & 4.072 \\
MGSD-LLAP-DAU & 1.446 & 1.520 & 1.693 & 2.114 \\ \bhline{1.1pt}
\end{tabular}
\end{table}

\subsubsection{COMPARISON METHODS}
We compare the denoising performance of the proposed method with the following existing methods: 
\begin{itemize}
    \item Graph low-pass filter (GLPF): $h(\lambda)=1/(1+\tau_2\lambda)$ \cite{shumanEmergingFieldSignal2013},
    \item Diffusion with heat kernel (HD): $h(\lambda)= \exp(-\tau_1\lambda)$ \cite{zhang2008Grapha},
    \item Singular value decomposition with hard shrinkage (SVDS) \cite{gavishOptimalHardThreshold2014},
    \item Autoencoder (AE) \cite{vincentStackedDenoisingAutoencoders2010},
    \item Graph convolutional network (GCN) \cite{DBLP:conf/iclr/KipfW17},
    \item Graph signal denoising via twofold graph smoothness regularization with DAU (TGSR-DAU) \cite{nagahama2022Multimodal}.
\end{itemize}
The first three methods are model-based approaches, the fourth and fifth ones are deep learning-based ones, and the last is a DAU-based one.

In this paper, we assume that graphs are not given a priori. 
Therefore, we construct (spatial and modality) graphs using the Gaussian RBF kernel \cite{keerthiAsymptoticBehaviorsSupport2003} for HD, GLPF, GCN, and TGSR-DAU. 
Note that HD, GLPF, and GCN can perform only on $\mathcal{G}_s$, i.e., they do not consider the modality direction.
The parameters used in model-based methods (HD, GLPF, and SVDS) are fitted with Bayesian parameter estimation \cite{bullConvergenceRatesEfficient2011}.
AE, GCN, TGSR-DAU, and the proposed method are trained the weights and the parameters from the training data, and the layers of the models are fitted $T=9$. 
Deep learning- and DAU-based methods are trained for 30 epochs using Adam optimizer \cite{DBLP:journals/corr/KingmaB14} whose learning rate is set to 0.01.
We use average of root mean square error (RMSE) as evaluation metrics.

\subsection{Denoising Results of Synthetic Dataset}\label{subsection:exp:syn}
\subsubsection{Quantitative and Qualitative Results}
The denoising results are presented in Table \ref{table:rmse_Synthetic}.
As shown in the table, the proposed method maintains low RMSEs under all noise situations. 
This indicates that leveraging the inter-modality relationship through the simultaneous graph learning process is effective for denoising of multimodal graph signals.

The proposed method shows a robust performance with various noise strengths compared to GLPF and TGSR-DAU. 
This is because GLPF and TGSR-DAU directly use noisy signals for graph construction, making them noise-sensitive. 
In contrast, twofold graph learning phases in our method can handle noisy signals well. 
HD, AE, and GCN show similar reconstruction errors across different noise levels, suggesting over-smoothing along the spatial direction (as shown in Fig. \ref{fig:visualization_syn}). 

The visualization results are shown in Fig. \ref{fig:visualization_syn}.
Model-based methods such as GLPF, HD, and SVDS tend to have limited denoising performances or oversmoothing has occurred.
AE and GCN, which are deep learning-based methods, extract clusters in the spatial domain, but the denoised signals have a low dynamic range.
In contrast, the proposed method has high denoising performance due to the sequential learning of spatial and modality graphs.

\subsubsection{Learned Graphs}
The adjacency matrices of the graphs estimated by the proposed method are shown in Figs. \ref{fig:Ws_syn} and \ref{fig:Wm_syn}. 
The figures show the obtained graphs in a layer-by-layer manner.
Basically, trained spatial graphs in the early layers (shown in Fig. \ref{fig:Ws_syn}(b)--(f)) have small edge weights: This implies the proposed method performs global denoising in the early stage.
The edge weights are getting larger when the layer goes deeper.
This could result in local denoising close to the output.
The spatial graph at the last layer, Fig. \ref{fig:Ws_syn}(j), is very sparse which suggests that eight layers for the spatial signal denoising could be enough.

\subsubsection{Learned Parameters} \label{subsection:ablation-learned-parameters}
Figure \ref{fig:learned_parameters} plots the values of the learned parameters $\{\alpha_s^{(t)}, \beta_s^{(t)}, \gamma_s^{(t)}, \alpha_m^{(t)}, \beta_m^{(t)}, \gamma_m^{(t)}\}$ against layers.
As shown below, the learned parameters exhibit different trends depending on their roles:

\textbf{Graph learning: }
Parameters $\beta_s$, $\beta_m$, $\gamma_s$ and $\gamma_m$ tend to slightly decrease when the layer $t$ increases. 
The large $\beta_e$ ($e \in \{s,m\}$) and $\gamma_e$ ($e \in \{s,m\}$) promote node connectivity, and sparseness of edge weights, respectively. 
In the early layers, large parameters promote sparse topologies with weak edge weights. 
In the later layers, in contrast, graphs become denser with stronger weights. 
This is because the signal is denoised in the later layers, i.e., it is more reliable than that in the early layers.
It is beneficial for graph learning.

\textbf{Graph signal denoising: }
As can be seen in Fig. \ref{fig:learned_parameters}, $\alpha_s$ tends to decrease and $\alpha_m$ tends to increase while both tendencies are slight.

The large $\alpha_e$ ($e \in \{s,m\}$) promotes a strong denoising and the small one corresponds to a weak denoising.
In this sense, the spatial and modality domain denoising processes work complementarily: In the earlier layers, stronger denoising is performed in the spatial domain signals, and in the later layers, that in the modality domain is (slightly) dominant.

This would be due to the signal characteristics used in this experiment:
Since we use the piecewise constant signals along the modality domain which does not follow our smoothness assumption, the strong denoising for the modality domain is only performed in the later layers that have a more reliable modality graph than the earlier ones.

\subsubsection{Effectiveness of Joint Graph Learning and Denoising} \label{subsection:ablation-gl-and-gd}
To verify the effectiveness of our joint graph learning and denoising strategy, we conduct an ablation study comparing the proposed method with a baseline that uses fixed graph.
Table \ref{table:ablation_rmse} summarizes the results, where the compared methods are defined as follows:
\begin{itemize}
    \item \textit{Oracle}: Denoising with the ground-truth graphs ($\mathbf{L}_s$ and $\mathbf{L}_m$).
    \item \textit{Learned ($\mathbf{L}_s$ and $\mathbf{L}_m$)}: Denoising with learned graphs using an RBF kernel.
    \item \textit{Learned ($\mathbf{L}_s$)}: Denoising with a fixed RBF-based $\mathbf{L}_s$ and a learned $\mathbf{L}_m$ in our algorithm.
    \item \textit{Learned ($\mathbf{L}_m$)}: Denoising with a fixed RBF-based $\mathbf{L}_m$ and a learned $\mathbf{L}_s$ in our algorithm.
    \item \textit{Proposed}: Our full model with simultaneous learning of both graphs.
\end{itemize}

The proposed method outperforms all comparative baselines.
Interestingly, \textit{Proposed} even outperforms \textit{Oracle}.
This implies that, in this experiment, the layer-wise graph learning could yield graphs better suited for signal denoising than the ground-truth ones. 
Furthermore, RMSEs of \textit{Proposed} are significantly better than the other fixed graph-based methods.
This further highlights the effectiveness of the proposed method.

\subsubsection{Empirical Convergence Analysis}
The proposed framework does not guarantee the global convergence of problem \eqref{equation:objective_proposed}. 
Furthermore, it is difficult to properly evaluate the trajectory of the objective function values of \eqref{equation:objective_proposed} across iterations, since the hyperparameters are learned and vary across layers. 
Instead, we evaluate the RMSE of the denoised signals output from each layer to analyze the behavior of the iterative process.

Figure \ref{fig:rmse_vs_layer} shows the layer-wise RMSEs for various noise levels. 
As can be observed, the RMSEs are (almost consistently) not increasing
as the layer number increases.
Therefore, the internal alternating optimization is (empirically) expected to be stable and not to diverge.
\color{black}

\begin{figure*}[t]
    \begin{minipage}[t]{0.30\linewidth}
        \centering
        \includegraphics[width=5cm]{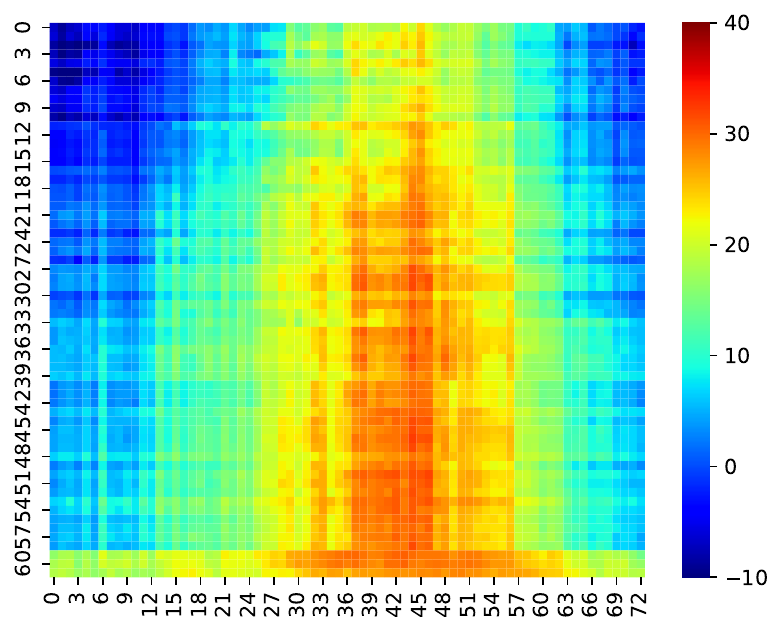}
        \subcaption{Original}
    \end{minipage}%
    \hfill
    \begin{minipage}[t]{0.30\linewidth}
        \centering
        \includegraphics[width=5cm]{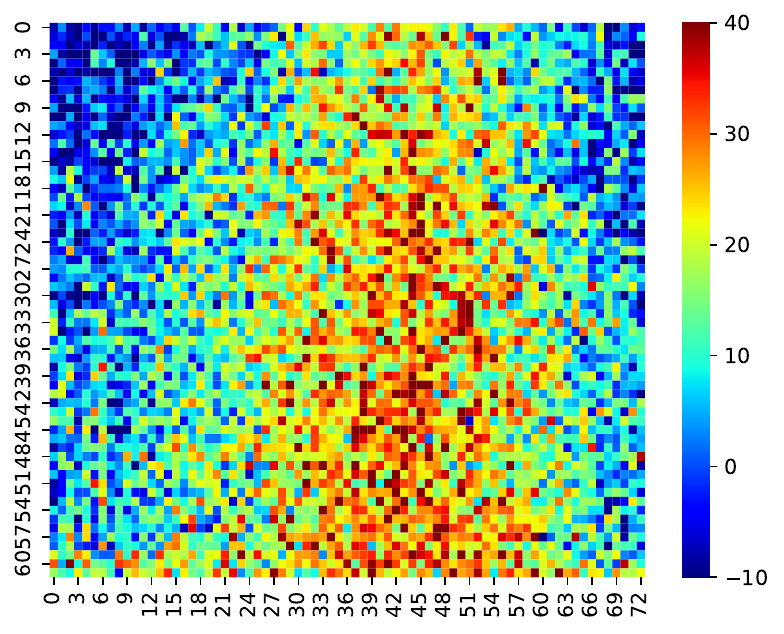}
        \subcaption{Observed (9.00) }
    \end{minipage}
    \hfill
    \begin{minipage}[t]{0.30\linewidth}
        \centering
        \includegraphics[width=5cm]{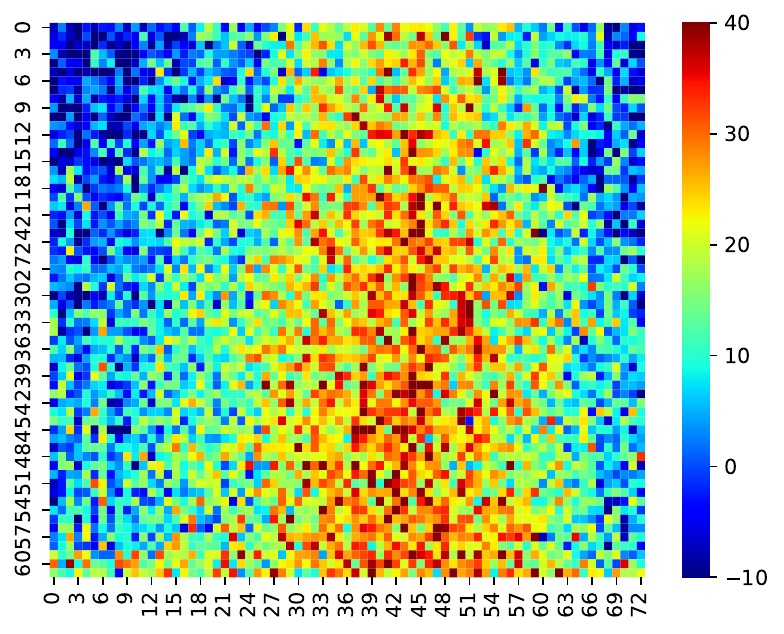}
        \subcaption{GLPF (8.00)}
    \end{minipage}
    \hfill
    \begin{minipage}[t]{0.30\linewidth}
        \centering
        \includegraphics[width=5cm]{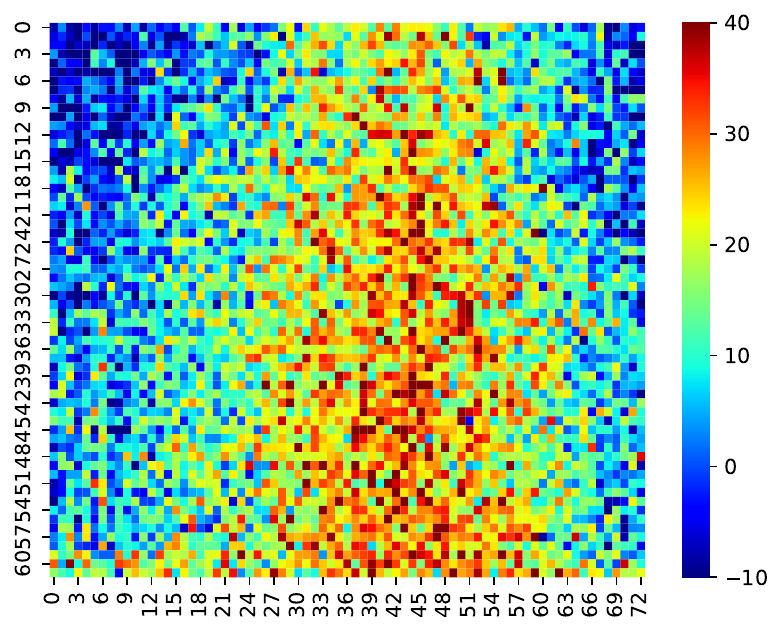}
        \subcaption{HD (8.53)}
    \end{minipage}
    \hfill
    \begin{minipage}[t]{0.30\linewidth}
        \centering
        \includegraphics[width=5cm]{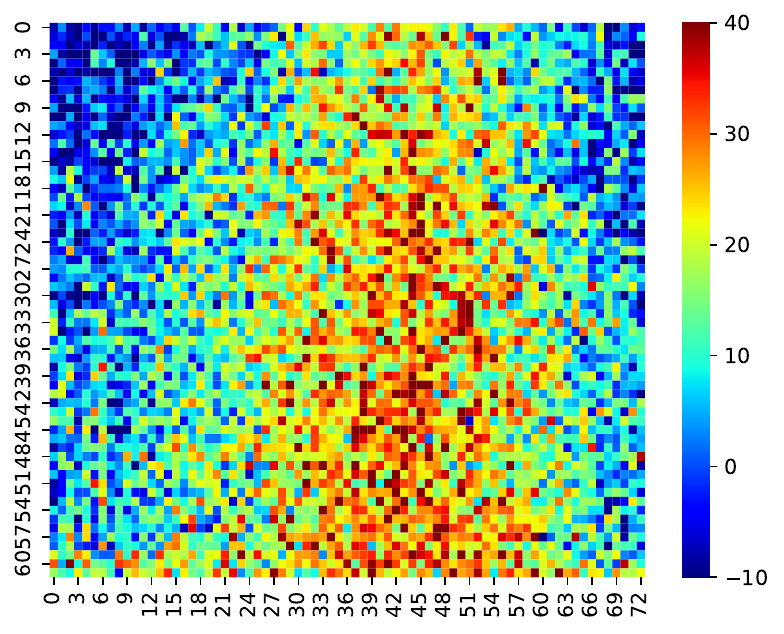}
        \subcaption{SVDS (8.95)}
    \end{minipage}
    \hfill
    \begin{minipage}[t]{0.30\linewidth}
        \centering
        \includegraphics[width=5cm]{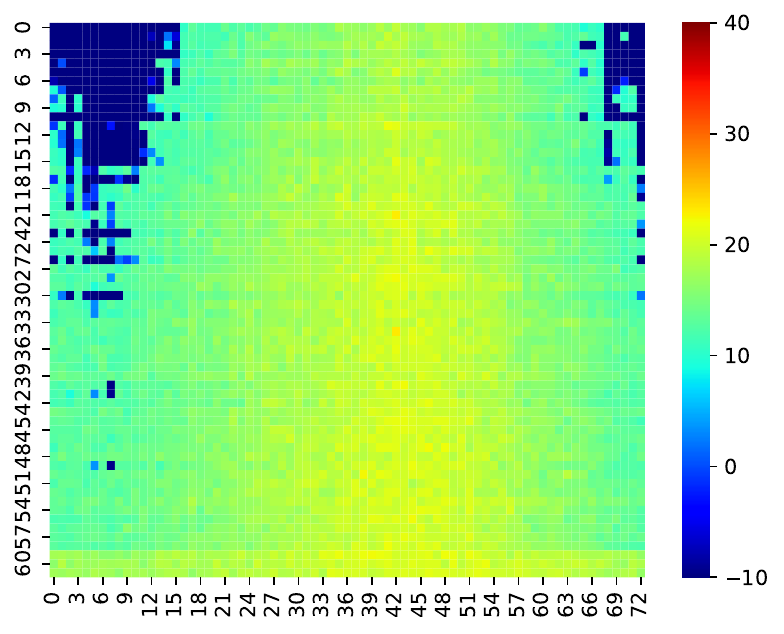}
        \subcaption{AE (5.92)}
    \end{minipage}
    \hfill
    \begin{minipage}[t]{0.30\linewidth}
        \centering
        \includegraphics[width=5cm]{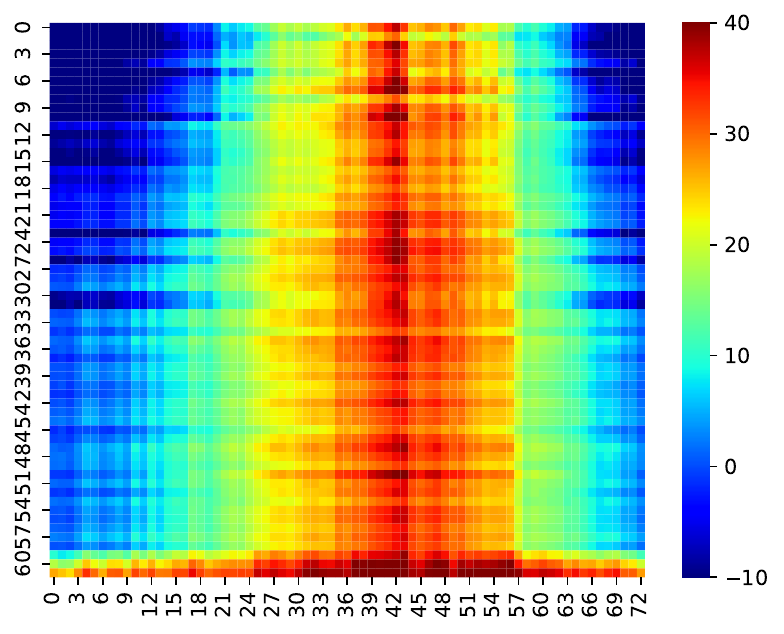}
        \subcaption{GCN (5.19)}
    \end{minipage}
    \hfill
    \begin{minipage}[t]{0.30\linewidth}
        \centering
        \includegraphics[width=5cm]{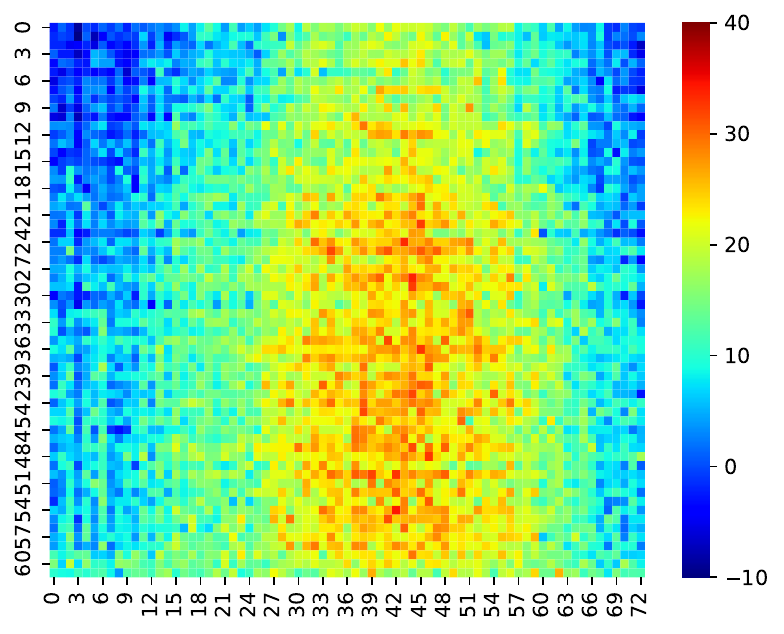}
        \subcaption{TGSR-DAU (4.19)}
    \end{minipage}
    \hfill
    \begin{minipage}[t]{0.30\linewidth}
        \centering
        \includegraphics[width=5cm]{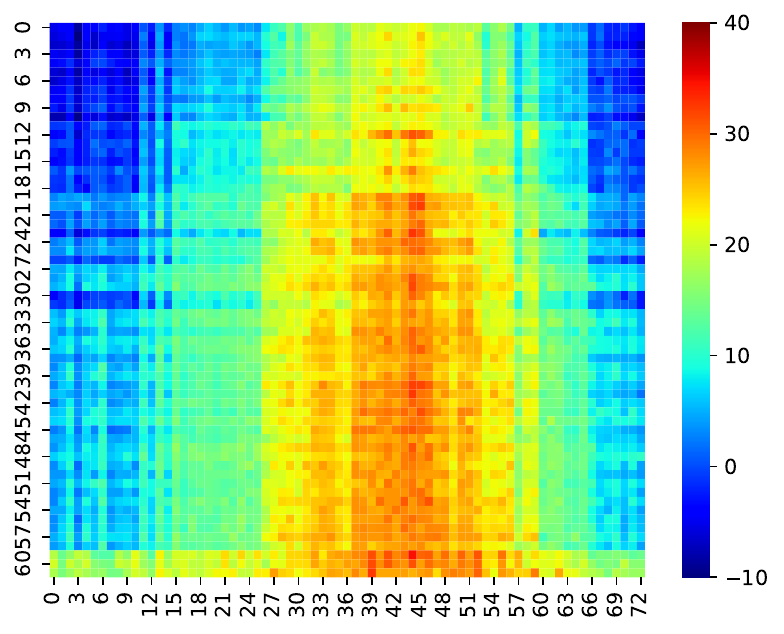}
        \subcaption{MGSD-LLAP-DAU (2.31)}
    \end{minipage}
    \caption{Denoising results for real-world dataset ($\sigma=9$). The vertical indices correspond to the spatial graph numbers and the horizontal ones are the modality. 
        Parenthesized values in subcaptions are RMSE.
        }
    \label{fig:visualization_real}
\end{figure*}

\begin{figure*}[t]
    \begin{minipage}[t]{0.2\linewidth}
        \centering
        \includegraphics[width=3.5cm]{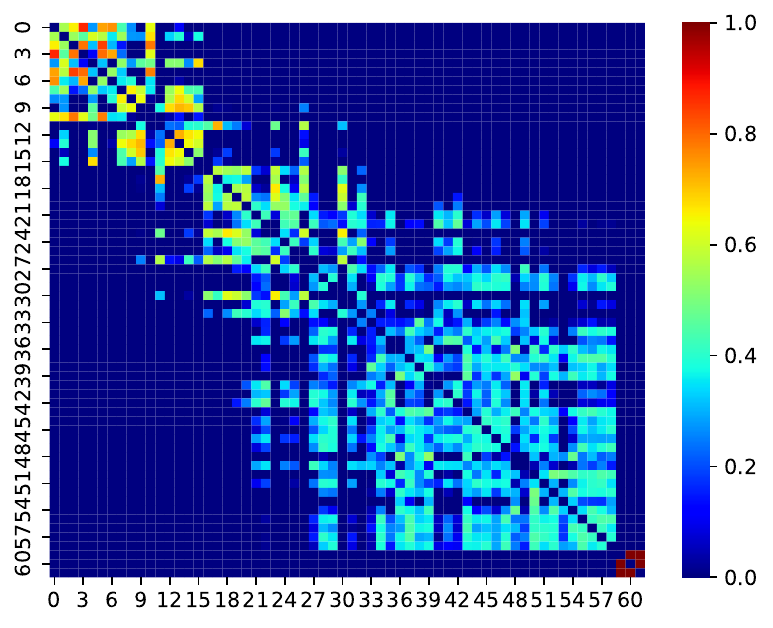}
        \subcaption{Layer 1}
    \end{minipage}%
    \hfill
    \begin{minipage}[t]{0.2\linewidth}
        \centering
        \includegraphics[width=3.5cm]{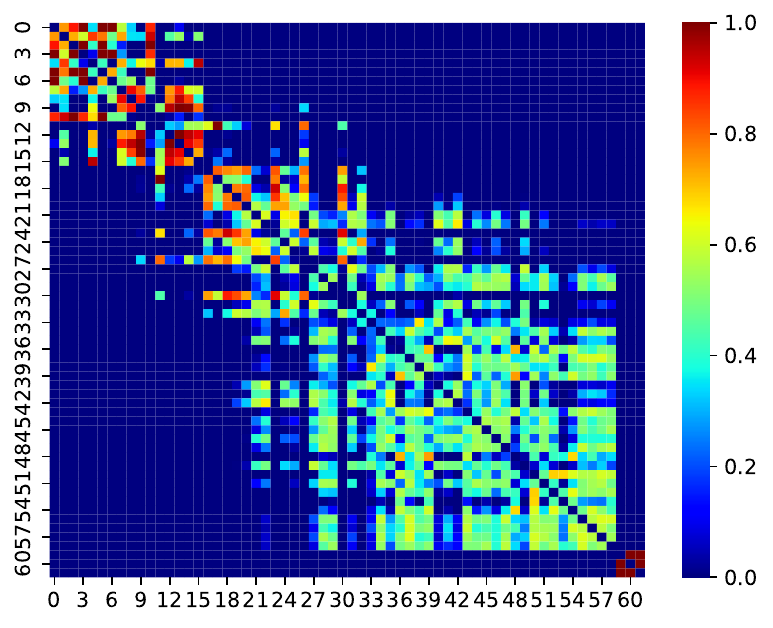}
        \subcaption{Layer 2}
    \end{minipage}%
    \hfill
    \begin{minipage}[t]{0.2\linewidth}
        \centering
        \includegraphics[width=3.5cm]{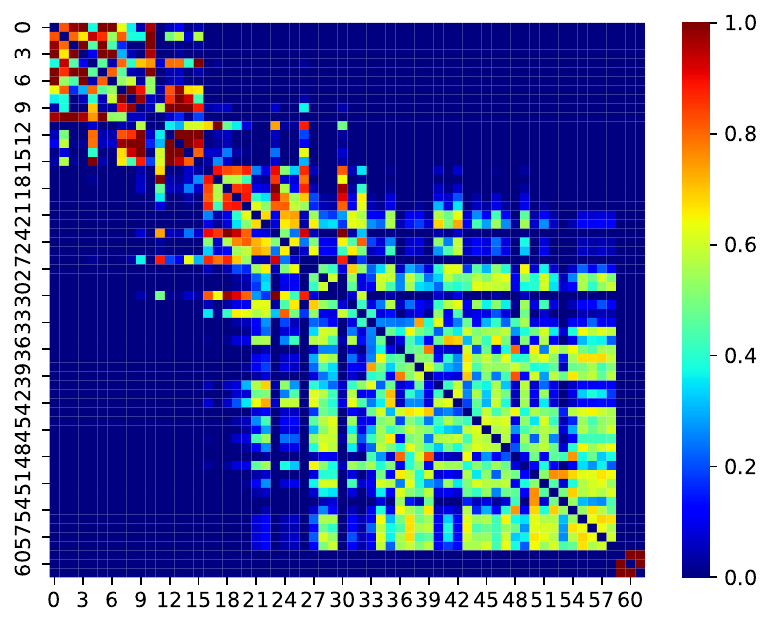}
        \subcaption{Layer 3}
    \end{minipage}%
    \hfill
    \begin{minipage}[t]{0.2\linewidth}
        \centering
        \includegraphics[width=3.5cm]{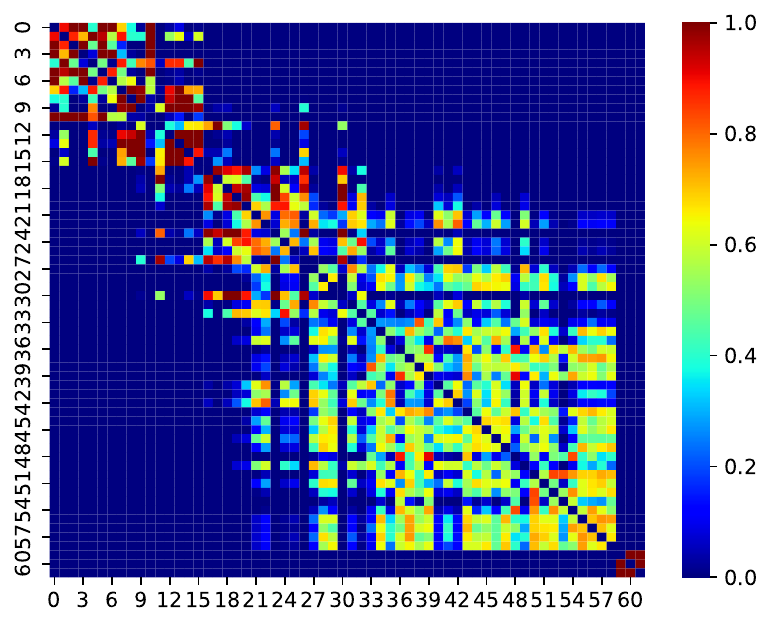}
        \subcaption{Layer 4}
    \end{minipage}%
    \hfill
    \begin{minipage}[t]{0.2\linewidth}
        \centering
        \includegraphics[width=3.5cm]{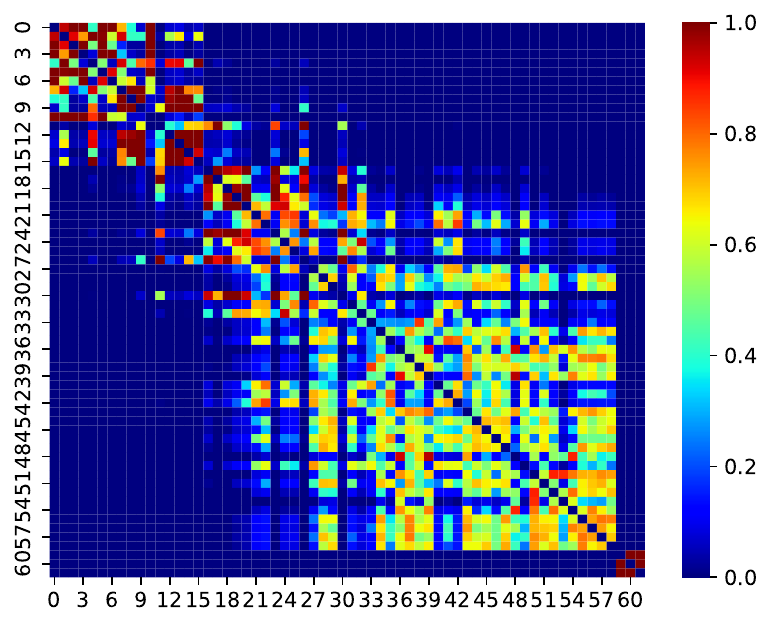}
        \subcaption{Layer 5}
    \end{minipage}%
    \hfill
    \begin{minipage}[t]{0.2\linewidth}
        \centering
        \includegraphics[width=3.5cm]{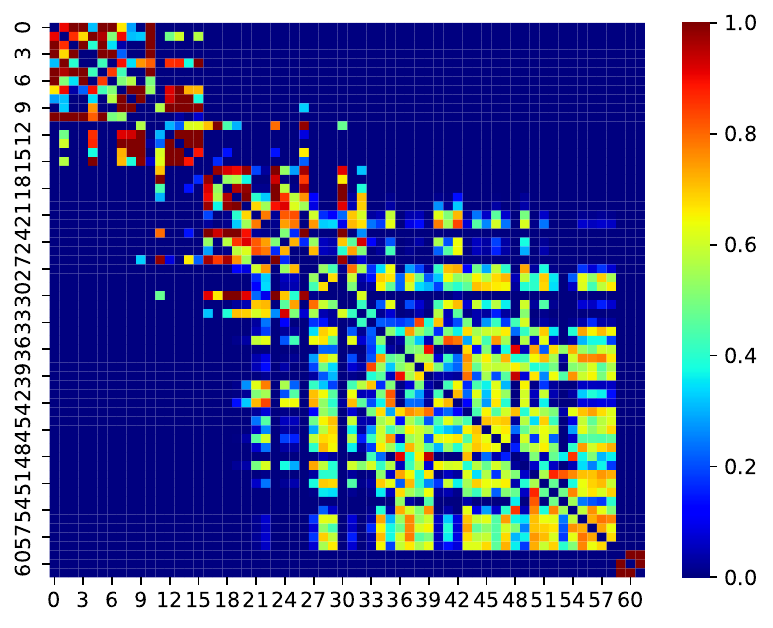}
        \subcaption{Layer 6}
    \end{minipage}%
    \begin{minipage}[t]{0.2\linewidth}
        \centering
        \includegraphics[width=3.5cm]{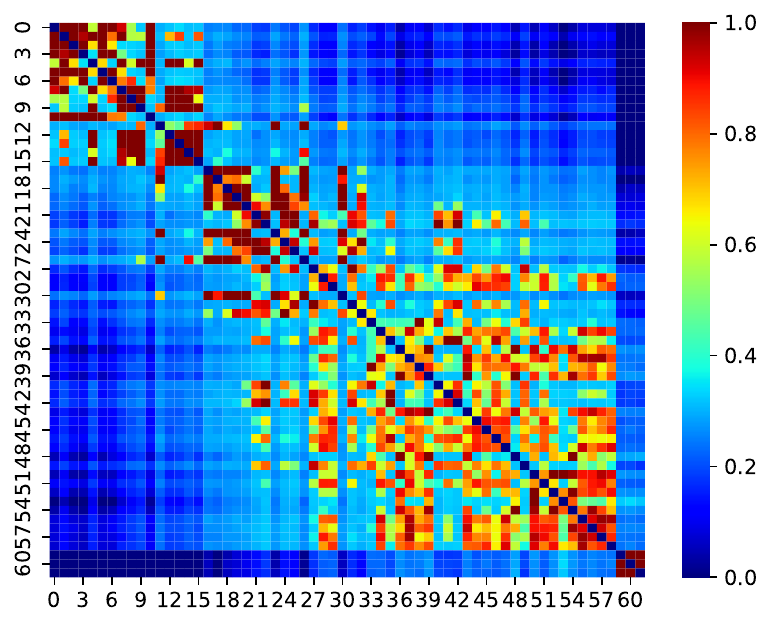}
        \subcaption{Layer 7}
    \end{minipage}%
    \begin{minipage}[t]{0.2\linewidth}
        \centering
        \includegraphics[width=3.5cm]{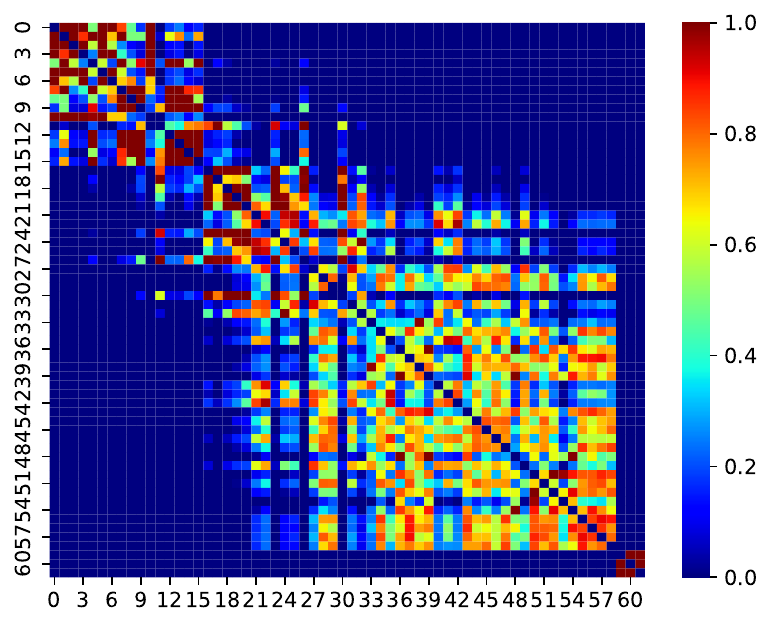}
        \subcaption{Layer 8}
    \end{minipage}%
    \begin{minipage}[t]{0.2\linewidth}
        \centering
        \includegraphics[width=3.5cm]{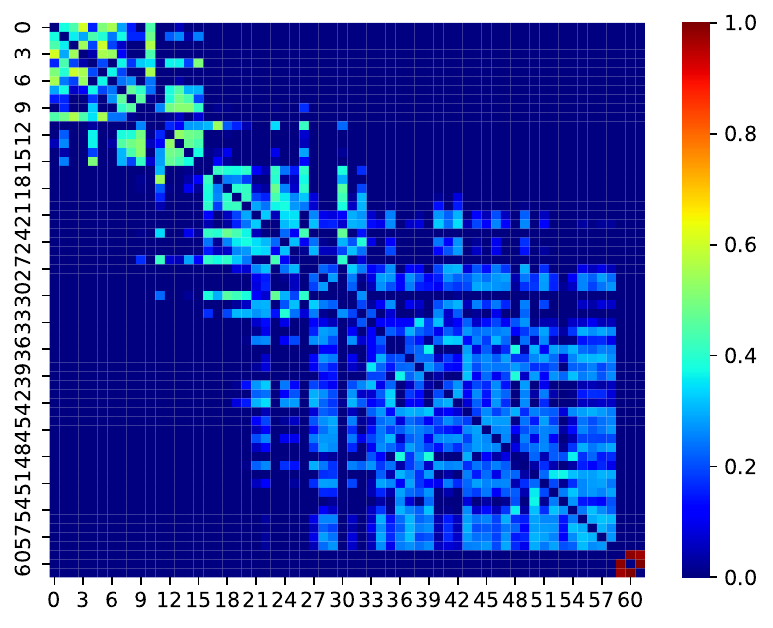}
        \subcaption{Layer 9}
    \end{minipage}%
    \hfill
    \caption{Visualization of learned spatial graph $\mathbf{W}_{s}$ for real-world dataset.
    }
    \label{fig:Ws_real}
\end{figure*}

\begin{figure*}[t]
    \begin{minipage}[t]{0.2\linewidth}
        \centering
        \includegraphics[width=3.5cm]{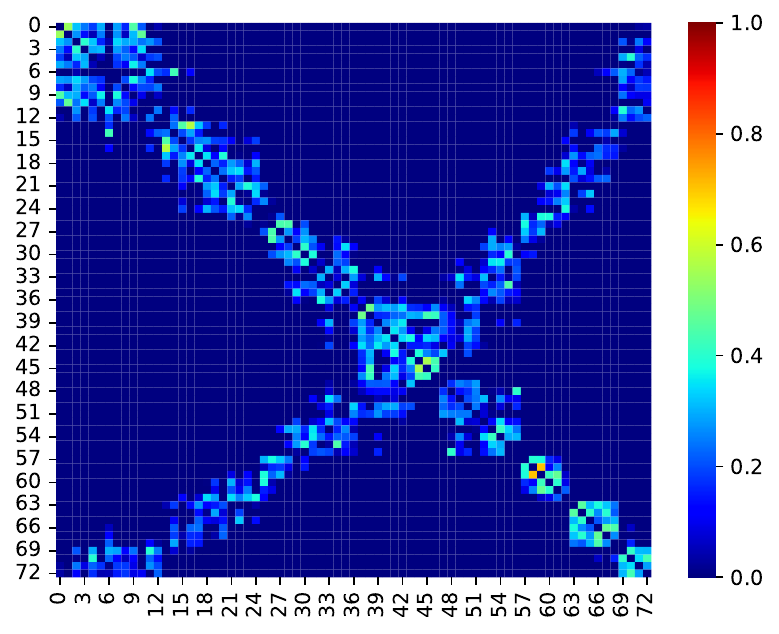}
        \subcaption{Layer 1}
    \end{minipage}%
    \hfill
    \begin{minipage}[t]{0.2\linewidth}
        \centering
        \includegraphics[width=3.5cm]{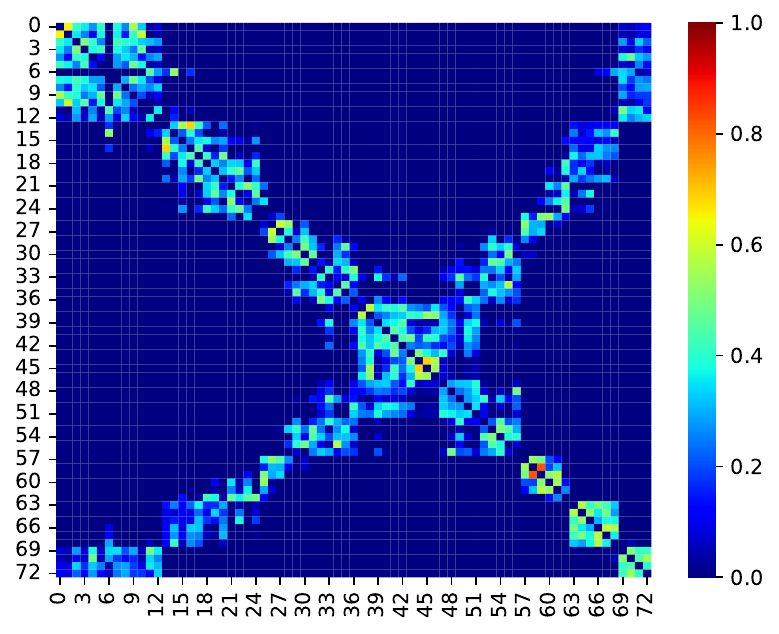}
        \subcaption{Layer 2}
    \end{minipage}%
    \hfill
    \begin{minipage}[t]{0.2\linewidth}
        \centering
        \includegraphics[width=3.5cm]{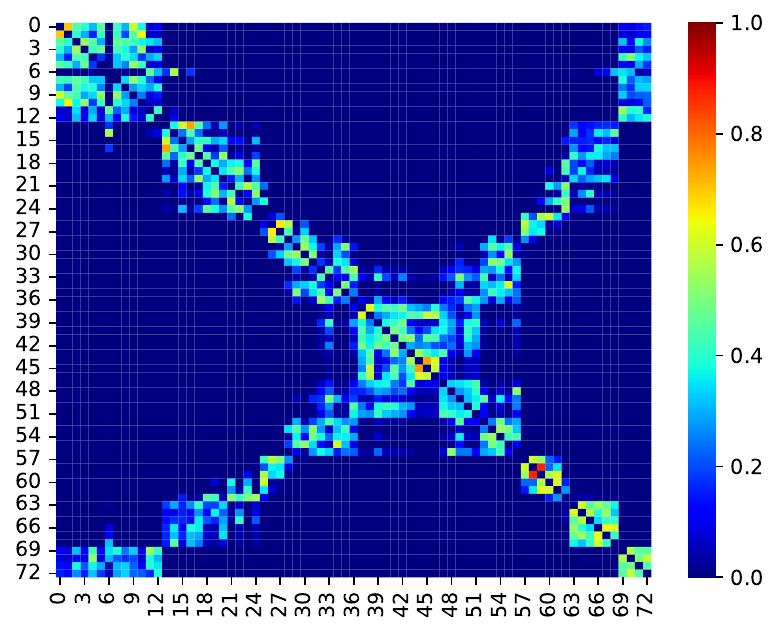}
        \subcaption{Layer 3}
    \end{minipage}%
    \hfill
    \begin{minipage}[t]{0.2\linewidth}
        \centering
        \includegraphics[width=3.5cm]{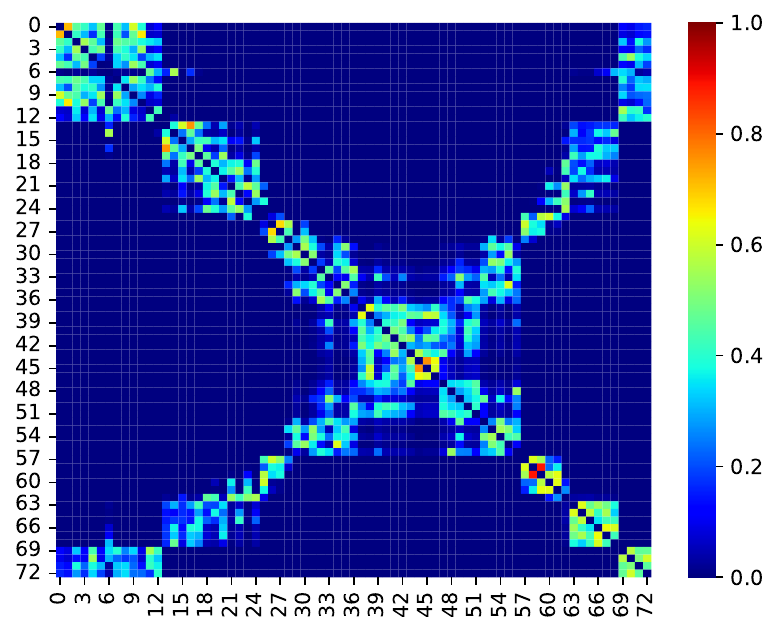}
        \subcaption{Layer 4}
    \end{minipage}%
    \hfill
    \begin{minipage}[t]{0.2\linewidth}
        \centering
        \includegraphics[width=3.5cm]{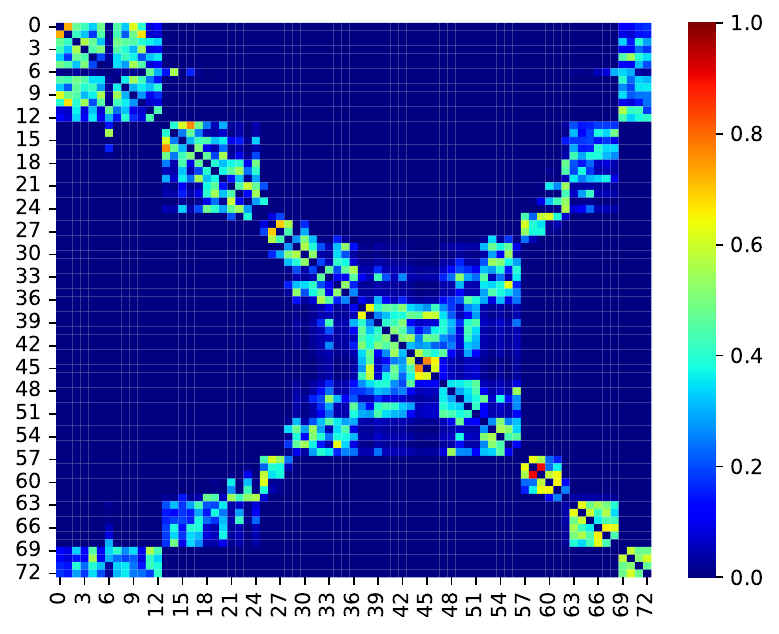}
        \subcaption{Layer 5}
    \end{minipage}%
    \hfill
    \begin{minipage}[t]{0.2\linewidth}
        \centering
        \includegraphics[width=3.5cm]{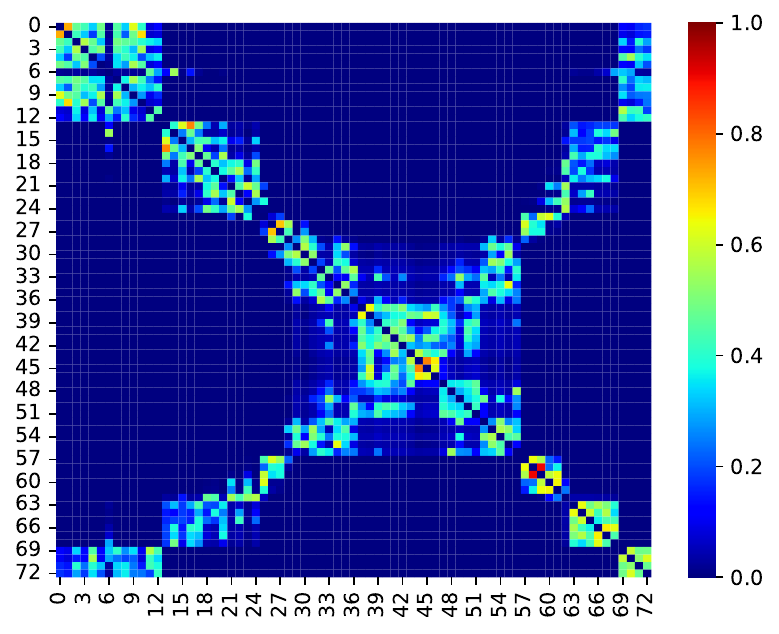}
        \subcaption{Layer 6}
    \end{minipage}%
    \begin{minipage}[t]{0.2\linewidth}
        \centering
        \includegraphics[width=3.5cm]{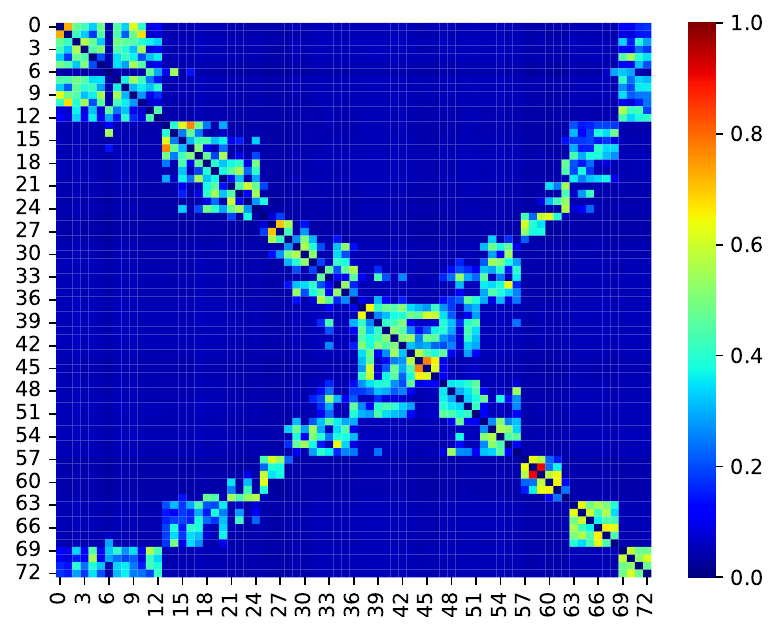}
        \subcaption{Layer 7}
    \end{minipage}%
    \begin{minipage}[t]{0.2\linewidth}
        \centering
        \includegraphics[width=3.5cm]{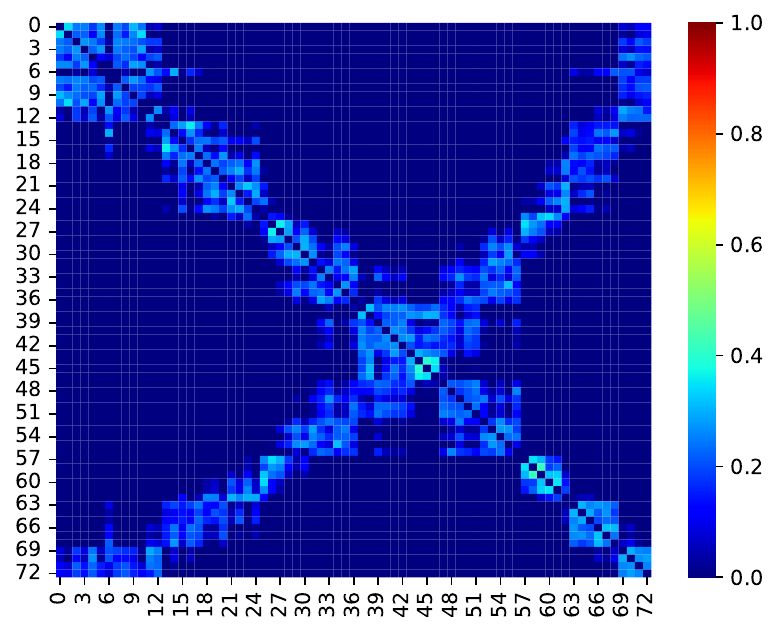}
        \subcaption{Layer 8}
    \end{minipage}%
    \begin{minipage}[t]{0.2\linewidth}
        \centering
        \includegraphics[width=3.5cm]{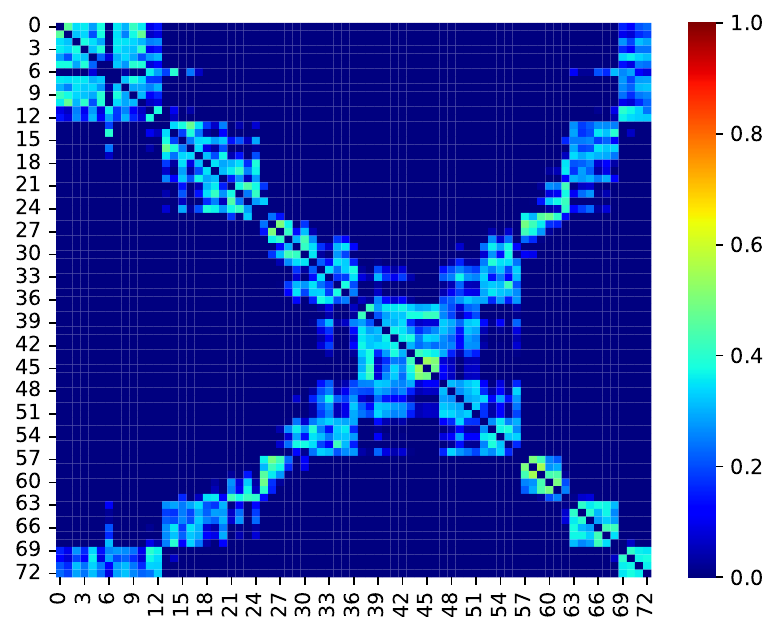}
        \subcaption{Layer 9}
    \end{minipage}%
    \hfill
    \caption{Visualization of learned modality graph $\mathbf{W}_{m}$ for real-world dataset.
    }
    \label{fig:Wm_real}
\end{figure*}

\subsection{Denoising Results of Real-world Dataset}
\subsubsection{Quantitative and Qualitative Results}
Table \ref{table:rmse_Weather} shows the denoising results for the real-world temperature dataset. 
Similar to the synthetic signals, the proposed method shows smaller RMSE than the alternative methods. 
Note that, for a graph signal denoising task, model-based methods often perform better than the deep learning-based ones, especially for small noise cases.
For large noises, deep learning-based methods are better than the model-based methods.
However, the predefined graphs limit the performance: It could be solved by our approach.
Note that AE may overfit to the data and produce the same outputs
regardless of the input. 

The visualization results are shown in Fig. \ref{fig:visualization_real}.
As the results of the synthetic dataset, GLPF, HD, and SVDS are not well performed and the resulting signals are quite similar to the noisy signal.
AE and GCN oversmoothed and overenhanced the multimodal signal, respectively.
TGSR-DAU can denoise the signal partially but noises still remain.
As observed, the proposed method is able to denoise the multimodal signal using both temporal and spatial information.

\subsubsection{Learned Graphs}
The adjacency matrices of the graphs estimated by the proposed method are shown in Figs. \ref{fig:Ws_real} and \ref{fig:Wm_real}.
Similar to the synthetic data, the graphs obtained in the first layer have low edge weights which results in global denoising.
In the following layers, edge weights are almost consistent until the seventh or eighth layers: These layers could correspond to local denoising.

In the spatial graph shown in Fig. \ref{fig:Ws_real}, nodes having close indices are likely to be connected that reflects geographic similarities: The station (i.e., node) indices are assigned from north to south.
Several softly-separated clusters are also observed.
This corresponds to similar climate regions in Japan: From top left to bottom right, clusters indicate the following regions: Hokkaido/Tohoku, Kanto, Chubu/Western Japan, and other isolated islands (very small cluster at the bottom right).

In the modality graph shown in Fig. \ref{fig:Wm_real}, edges are drawn between nodes with similar indices, suggesting a positive correlation between signals from nearby dates. 
Furthermore, nodes (i.e., dates) having similar climates are also connected.
For example, nodes 0 to 12 (early January to early March) are connected to nodes 66 to 72 (late November to late December), which reflects seasonal behavior of climates in Japan.

\section{CONCLUSION}
\label{section:conclusion}
This paper presents a simultaneous method of multimodal signal denoising on a twofold graph and graph learning using DAU.
The proposed algorithm alternates graph learning and signal denoising both for the spatial and modality domains.
We solve the iterative optimization by DAU where the parameters in the iterations are automatically tuned with training data.
The numerical results demonstrate that the proposed method is effective for denoising multimodal signals compared to conventional methods.

\appendix[Derivation of Iterative Algorithm for \eqref{equation:objective_function_sub_vech_indicator}]
PDS is an algorithm designed to solve a minimization problem written as follows:
\begin{equation}
    \min_{\bm{\ell}} g_{0}(\bm{\ell}) + g_{1}(\bm{\Psi}\bm{\ell}) + g_{2}(\bm{\ell}),
    \label{equation:pds_format}
\end{equation}
where $g_{0}$ and $g_{1}$ are proximable\footnote{Let $g \colon \mathbb{R}^{N} \rightarrow \mathbb{R} \cup \{ \infty \}$ be a proper lower semicontiunous convex function. The proximal operator $\textrm{prox}_{\mu g} \colon \mathbb{R}^{N} \rightarrow \mathbb{R}^{N}$ of $g$ with a parameter $\theta>0$ is defined by $\textrm{prox}_{\theta g} (\mathbf{x}) = {\argmin}_{\mathbf{y}} \{g(\mathbf{y}) + \frac{1}{2 \theta} \| \mathbf{x-y} \|_{2}^{2} \}$. If $\textrm{prox}_{\theta g}$ can be computed efficiently, the function $g$ is called proximable.} proper lower semicontinuous convex functions, and $g_{2}$ is a differentiable convex function whose gradient $\nabla g_{3}$ has a Lipschitz constant.
Applying \eqref{equation:pds_format} to our problem, each term in \eqref{equation:objective_function_sub_vech_indicator} can be assigned as follows:
\begin{equation}
    \begin{aligned}
        g_{0}(\bm{\ell}_{e})
        &= 
        \iota_{\cdot \leq \mathbf{0}}(\bm{\ell}_{e}), \\
        g_{1}( \bm{\Psi}_{e} \bm{\ell}_{e} ) 
        &=
        - \beta_{e} \mathbf{1}_{N_{e}}^{\top} \log(\bm{\Psi}_{e} \bm{\ell}_{e}), \\
        g_{2}(\bm{\ell}_{e}) 
        &= 
        \alpha_{e} \textrm{vec}(\mathbf{X}_{e}\mathbf{X}_{e}^{\top})^{\top} \bm{\Phi}_{e}\bm{\ell}_{e}
        +
        \gamma_{e} \| \bm{\ell}_{e} \|_{2}^{2}.  
    \end{aligned}
    \label{equation:pds_form_e}
\end{equation}
The proximal operator of $g_{0}$ and $g_{1}$, as well as $\nabla g_{2}$ are given by
\begin{equation}
    \begin{aligned}
        \textrm{prox}_{\theta_{e} g_{0}}(\ell) &= \min \{\ell, 0\}, \\
        \textrm{prox}_{\theta_{e} g_{1}}(\ell) &= \frac{\ell + \sqrt{\ell^{2} + 4 \beta_{e} \theta_{e}}}{2}, \\
        \nabla g_{2}(\bm{\ell}) &= \alpha_{e} \bm{\Phi}_{e}^{\top} \textrm{vec}(\mathbf{X}_{e} \mathbf{X}_{e}^{\top}) + 2 \gamma_{e} \bm{\ell},
    \end{aligned}
\end{equation}
where $\theta_{e} \in (0, +\infty)$ denotes the stepsize (refer to \cite{komodakisPlayingDualityOverview2015a} for the conditions of algotrithm convergence).

\bibliographystyle{IEEEtran}
\bibliography{multimodal}

\vfill\pagebreak

\vfill

\end{document}